\documentclass[showpacs,10pt,aps,prd,reprint,onecolumn,notitlepage,nofootinbib,superscriptaddress]{revtex4-1}

\usepackage{hyperref}
\usepackage{amsmath}
\usepackage{mathtools}
\usepackage{mathrsfs}
\usepackage{bm}
\usepackage{amssymb}
\usepackage{graphicx}
\usepackage[]{subfigure}
\usepackage{filecontents}
\usepackage[dvipsnames]{xcolor}

\hypersetup{
    bookmarks=true,         
    pdftoolbar=true,        
    pdfmenubar=true,        
    pdffitwindow=true,     
    pdfstartview={FitH},     
    pdftitle={My title},     
    pdfauthor={author},      
    pdfsubject={Subject},    
    pdfcreator={Creator},    
    pdfproducer={Producer},  
    pdfkeywords={keyword1} 
    pdfnewwindow=true,       
    colorlinks=true ,        
    linkcolor=blue  ,        
    citecolor=blue,      
    filecolor=green,       
    urlcolor=blue            
}

\newcommand{\ed}{\mathrm{d}}
\newcommand{\mr}{\mathrm{r}}
\newcommand{\tL}{\tilde{\mathcal{L}}}
\newcommand{\tE}{\tilde{\mathcal{E}}}
\newcommand{\K}{{\mathcal{K}}}
\newcommand{\tb}{{\tilde{b}}}
\newcommand{\teta}{{\tilde{\eta}}}
\makeatletter
\DeclareFontFamily{U}{tipa}{}
\DeclareFontShape{U}{tipa}{m}{n}{<->tipa10}{}
\newcommand{\arc@char}{{\usefont{U}{tipa}{m}{n}\symbol{62}}}%

\newcommand{\arc}[1]{\mathpalette\arc@arc{#1}}

\newcommand{\arc@arc}[2]{%
  \sbox0{$\m@th#1#2$}%
  \vbox{
    \hbox{\resizebox{\wd0}{\height}{\arc@char}}
    \nointerlineskip
    \box0
  }%
}
\makeatother

\begin{document}

\title{Ergosphere, photon region structure, and the shadow of a rotating charged Weyl black hole}

\author{Mohsen Fathi}
\email{mohsen.fathi@postgrado.uv.cl}
\affiliation{Instituto de F\'isica y Astronom\'ia, Universidad de Valpara\'iso, Avenida Gran Breta\~na 1111, Valpara\'iso, Chile}

\author{Marco Olivares}
\email{marco.olivaresr@mail.udp.cl}
\affiliation{Facultad de
              Ingenier\'{i}a y Ciencias, Universidad Diego Portales, Avenida Ej\'{e}rcito
              Libertador 441, Casilla 298-V, Santiago, Chile}

\author{J.R. Villanueva}
\email{jose.villanueva@uv.cl}
\affiliation{Instituto de F\'isica y Astronom\'ia, Universidad de Valpara\'iso, Avenida Gran Breta\~na 1111, Valpara\'iso, Chile}

\date{\today}

\begin{abstract}
In this paper we explore the photon region and the shadow of the rotating counterpart of a static charged Weyl black hole, which has been previously discussed according to null and time-like geodesics. The rotating black hole shows strong sensitivity to the electric charge and the spin parameter, and its shadow changes from being oblate to being sharp by increase in the spin parameter. {Comparing the calculated vertical angular diameter of the shadow with that of M87*, it turns out that the latter may possess about $10^{36}$ protons as its source of electric charge, if it is a rotating charged Weyl black hole}. A complete derivation of the ergosphere and the static limit is also presented.

\bigskip

\noindent{\textit{keywords}}: Weyl gravity, black hole shadow, ergosphere
\end{abstract}


\maketitle

\section{Introduction}

The recent black hole imaging of the shadow of M87*, done by the Event Horizon Telescope (EHT)  \cite{Akiyama:2019}, was another significant affirmation of general relativity. Despite this, the quest for finding reliable modified and alternative theories of gravity still remains in force. This mostly stems in modern galactic and extra-galactic evidences \cite{Rubin1980,Massey2010,Riess:1998,Perlmutter:1999,Astier:2012}, that have led to the consideration of dark matter and dark energy scenarios. On the other hand, since these exotic cosmological constituents are extremely hard to detect, some scientists believe that uplift in our knowledge about the action of the gravitational field, can exclude the necessity of dark matter and dark energy scenarios in cosmological descriptions. This is pursued through finding newer gravitational actions and studying the resultant theories \cite{Clifton2012pr}. The exploration of these theories, however, is not done only by means of cosmological observations. In fact, as in the general relativistic cases, the possible black hole solutions inferred from modified and alternative theories of gravity are constantly investigated regarding the photon trajectories in their exterior geometry, as well as their natural optical appearance to distant observers. This latter, because of the possible relevance between the shadow of black holes and their physical characteristics, has been set to rigorous numerical and analytical studies, in order to assess the black hole solutions regarding the observations. 

Beside the efforts done for recovering the dark matter and dark energy cosmological predictions, the selection of alternative theories of gravity is also based on the peculiarities of their gravitational action. {Along the same efforts, the fourth order theory of conformal gravity, has been appeared as one of the alternatives to compare the theoretical results with the cosmological and astrophysical observations, and some considerable constrains have been driven. This theory was formulated by H. Weyl in 1918 \cite{Weyl1918mz}, revived by Riegert in 1984 \cite{Riegert1984}, and since then, has been applied widely to explain the astrophysical phenomena (see for example Refs.~\cite{Mannheim:1997,Mannheim:2006,Diaferio:2011,Varieschi:2014b,Jizba:2015,Potapov:2016,Bambi:2017b,Zhang:2017,Zhou:2018}). It has also been argued that, unlike other cosmological models, this theory can avoid the age problem. The latter has been shown by testing the theory with the quasar APM 08279+5255 with the age of $(2.1 \pm 0.34)\times 10^{9}$ years at $z = 3.91$, indicating that the theory could accommodate the quasar within a 3$\sigma$  confidence level \cite{Yang:2013}. Furthermore, this theory has been applied to the determination of the energy carried by the gravitational waves \cite{Caprini:2018,Yang:2018,Momennia:2020,Faria:2020}.}

{Based on the above notes, in this paper, we consider a black hole solution which has been generated in the context of Weyl gravity.}  In fact, the black hole spacetimes inferred from theories of gravity are essentially constructed on the static vacuum solutions to their equations of motion. For the case of Weyl conformal gravity, such a solution was proposed by Mannheim and Kazanas in Ref.~\cite{Mannheim:1989}, where the authors connected the presence of flat galactic rotation curves, to an additional effective potential available in their solution. This way, Weyl gravity was proposed as an alternative to dark matter. Ever since, the theory has also been intended to recover the cosmological observations related to the dark energy scenario \cite{Mannheim:2005,Nesbet:2012}. Accordingly, and provided the Mannheim-Kazanas solution, the theory has been put to numerous investigations  \cite{Knox:1993fj,Edery:1997hu,Klemm:1998kf,Edery:2001at,Pireaux:2004id,Pireaux:2004xb,Diaferio:2008gh,Sultana:2010zz,Diaferio:2011kc,Mannheim:2011is,Tanhayi:2011dh,Said:2012xt,Lu:2012xu,Villanueva:2013Weyl,Mohseni:2016ylo,Horne:2016ajh,Lim:2016lqv,Varieschi2010,Hooft2010a,Hooft2010b,Hooft2011,Varieschi2012isrn,Varieschi2014gerg,Vega2014,Varieschi2014galaxies,Hooft2015,Deliduman:2015}. The black hole spacetime associated with this static solution, contains a cosmological term, and the rotating and electrically charged solutions to Weyl gravity have been also proposed in Ref.~\cite{Mannheim1991}, in the context of Kerr and Kerr-Newman spacetimes. 

In this work, we also consider a rotating charged black hole spacetime, inferred from Weyl gravity. This solution is indeed the rotating counterpart of a particular static charged black hole, introduced in Ref.~\cite{Payandeh:2012mj}. This charged Weyl black hole (CWBH) has been recently studied regarding the light propagation in vacuum and in plasmic medium \cite{Fathi:2020,Fathi:2020d}, and the motion of  neutral and charged particles  \cite{Fathi:2020b,Fathi:2020c}. As it is a rather interesting subject to understand how a black hole would appear in an observer's sky, it is therefore completely natural to discuss the optical properties of the rotating charged Weyl black hole (RCWBH) and its shadow. For the case of the Mannheim-Kazanas solution, a rotating version of the Weyl black hole has been given in Ref.~\cite{Varieschi:2014}, and its shadow has been discussed in Ref.~\cite{Mureika:2017}.

In this paper we follow these steps: In Sec.~\ref{sec:solution}, we introduce the CWBH solution and apply a modified Newman-Janis method to obtain its rotating counterpart (the RCWBH), and we discuss the horizon structure. In Sec.~\ref{sec:ergosphere}, this solution is studied in more details regarding the stationary and static observers, and accordingly, the ergosphere of the black hole is calculated and discussed. We use the first order light-like geodesic equations in Sec.~\ref{sec:shadow}, to proceed with studying the optical appearance of the black hole to distant observers, by means of calculating the photon region and the black hole shadow. A particular geometric method is then used to indicate the conformity between the deformation and the angular size of the shadow. We conclude in Sec.~\ref{sec:conlcusion}. Through the paper, we employ the geometric units, in which $G=c=1$.

\section{The black hole solution}\label{sec:solution}

The gravitational action corresponding to Weyl gravity is written in terms of the Weyl conformal tensor, as
\begin{equation}\label{eq:IWeyl}
    I_W=-\mathcal{K}\int{\ed^4x\sqrt{-g}\,\,C_{\alpha\beta\rho\lambda}C^{\alpha\beta\rho\lambda}},
 \end{equation}
in which  $\mathcal K$ is a coupling  constant, 
$g=\mathrm{det}(g_{\alpha\beta})$, and
\begin{equation}\label{eq:C}
C_{\alpha\beta\lambda\rho} = R_{\alpha\beta\lambda\rho}-\frac{1}{2}\left(g_{\alpha\lambda}
R_{\beta\rho}-g_{\alpha\rho}R_{\beta\lambda}-g_{\beta\lambda}R_{\alpha\rho}+g_{\beta\rho}R_{\alpha\lambda}\right)\\
+\frac{R}{6}\left(g_{\alpha\lambda}g_{\beta\rho}-g_{\alpha\rho}g_{\beta\lambda}\right).
\end{equation}
The conformal symmetry of this theory is inferred from the fact that the action $I_W$ is invariant under the conformal transformation $g_{\alpha\beta}(x) = e^{2 \alpha(x)} g_{\alpha\beta}(x)$, in which, $2 \alpha(x)$ is termed as the local spacetime stretching. Recasting $I_W$ in the form
\begin{equation}\label{eq:IWeyl-2}
    I_W=-\mathcal{K}\int
\ed^4x
\sqrt{-g}\,\left(R^{\alpha\beta\rho\lambda}R_{\alpha\beta\rho\lambda}-2R^{\alpha\beta}R_{\alpha\beta}+\frac{1}{3}R^2\right),
\end{equation}
it is seen that the Gauss-Bonnet term $\sqrt{-g}\,(R^{\alpha\beta\rho\lambda}R_{\alpha\beta\rho\lambda}-4R^{\alpha\beta}R_{\alpha\beta}+R^2)$
is a total divergence and can be disregarded. We can therefore simplify the action as \cite{Mannheim:1989,Kazanas:1991}
\begin{equation}\label{eq:IWeyl-3}
    I_{W}=-2\mathcal{K}\int{\ed^4x}\sqrt{-g}\,\,\left(R^{\alpha\beta}R_{\alpha\beta}-\frac{1}{3}R^2\right).
\end{equation}
The equations of motion are then obtained by applying the principle of least action, $\frac{\delta{I_W}}{\delta{g_{\alpha\beta}}} = 0$, providing the Bach equation 
\begin{eqnarray}\label{eq:Bach}
    W_{\alpha\beta} &=& \mathfrak{P}_{\gamma\delta}\,{{{C_{\alpha}}^{\gamma}}_{\beta}}^{\delta}+\Box\, \mathfrak{P}_{\alpha\beta}-\nabla ^{\gamma}\nabla _{\alpha}\mathfrak{P}_{\beta\gamma}\nonumber\\
    &=& \frac{1}{4\mathcal{K}} T_{\alpha\beta},
\end{eqnarray}
in which $W_{\alpha\beta}$ is the trace-less Bach tensor, $\Box \equiv \nabla^\gamma \nabla_\gamma$, and 
\begin{equation}
    \mathfrak{P}_{\alpha\beta}=\frac{1}{2}\left(R_{\alpha\beta}-\frac{1}{4} g_{\alpha\beta} R\right),
\end{equation}
is the  Schouten tensor, which is expressed in terms of the Ricci tensor and the scalar curvature of the spacetime. For a vanishing energy-momentum tensor ($T_{\alpha\beta}=0$), the Mannheim-Kazanas static spherically symmetric solution is given by the metric 
\begin{equation}
	{\rm d}s^{2}=-B(r)\, {\rm d}t^{2}+\frac{{\rm d}r^{2}}{B(r)}+r^{2}({\rm d}\theta^{2}+\sin^{2}\theta\,
	{\rm d}\phi^{2}) \label{metr}
\end{equation} 
in the usual Schwarzschild coordinates ($-\infty < t < \infty$, $r\geq0$, $0\leq\theta\leq\pi$ and $0\leq\phi\leq 2\pi$), with the lapse function \cite{Mannheim:1989}
\begin{equation}\label{eq:originalWeylB(r)}
 B(r) = 1-\frac{\zeta\left(2-3\zeta\rho\right)}{r}  - 3\zeta \rho + \rho r - \sigma r^2. 
\end{equation}
The three-dimensional integration constants $\zeta$, $\rho$ and $\sigma$, help recovering the general relativistic black hole spacetimes, in the way that  for $\rho\rightarrow 0$, the Schwarzschild-de Sitter solution is recovered. This corresponds to distances much smaller than $1/\rho$. 

{In the case that the energy-momentum tensor is associated with the electrostatic vector potential 
\begin{equation}\label{eq:vectorPotential}
    A_\alpha = \left(
    \frac{q}{r},0,0,0
    \right),
\end{equation}
for a spherically symmetric massive source of electric charge $q$, in Ref.~\cite{Payandeh:2012mj}, the vacuum Bach equation (i.e. $W_{\alpha\beta} = 0$ in Eq.~(\ref{eq:Bach})) has been solved for the reference lapse function
{
\begin{equation}\label{eq:Referencef}
    B(r) = 1-\frac{\mathrm{2\mathrm{C}}}{r}-\frac{1}{3}f(r),
\end{equation}
that provides the simple solution
\begin{equation}\label{eq:f(r)}
    f(r) = -\left(\frac{6 \mathrm{C}}{r}+c_1 r^2 +c_2 r\right),
\end{equation}
for which, substitution in Eq. \eqref{eq:Referencef} yields}
\begin{equation}\label{eq:F_metric_0}
    B(r) = 1+\frac{1}{3}\left(c_2 r + c_1 r^2\right).
\end{equation}
The determination of the coefficients $c_1$ and $c_2$ in the above lapse function, is done through the weak-field limit approach, by considering the last two terms of $B(r)$ as perturbations on the Minkowski spacetime \cite{Tanhayi:2011dh,Payandeh:2012mj}. This way, by taking into account a spherically symmetric source of mass $\tilde{m}$, charge $\tilde{q}$ and radius $\tilde{r}$, the Poisson equation $\nabla^2 h_{\alpha\beta} = 8\pi T_{\alpha\beta}$, with $h_{\alpha\beta} = g_{\alpha\beta}-\eta_{\alpha\beta}$, has the following 00 component in Minkowski spacetime:
\begin{equation}\label{eq:Poisson_1}
    \nabla^2 h_{00} \equiv -\frac{1}{3}\left(\frac{\partial^2}{\partial r^2} + \frac{2}{r}\frac{\partial}{\partial r}\right) \left({c_1 r^2}+{c_2 r}\right) = 8\pi T_{00} \\
    = 8\pi\left( \frac{\tilde{m}}{\frac{4}{3}\pi \tilde{r}^3}+\frac{1}{8\pi}\frac{\tilde{q}^2}{r^4}\right),
\end{equation}
in which, $T_{00}$ is the scalar part of the total energy-momentum tensor, that corresponds to the mass and charge of the source. From Eq.~\eqref{eq:Poisson_1}, we get to the relation \cite{Payandeh:2012mj}
{\begin{equation}\label{eq:c1c2}
    c_2 = -\left(\frac{9\tilde{m}\,r}{\tilde{r}^3}+\frac{3}{2}\frac{\tilde{q}^2}{r^3}+3c_1 r\right),
\end{equation}}
that provides}
\begin{equation}\label{lapse}
	B(r)=1-\frac{r^{2}}{\lambda^{2}}
	-\frac{Q^2}{4 r^2},
\end{equation}
where
\begin{eqnarray}
&&\frac{1}{\lambda^2}=\frac{3 \tilde{m}}{\tilde{r}^{3}}   +\frac{2 \tilde\varepsilon}{3},\label{par1}\\
&&Q=\sqrt{2}\, \tilde{q}.\label{par2}
\end{eqnarray}
{The coefficient $\tilde\varepsilon\equiv c_1$ in Eq.~\eqref{par1}, is intended to recover the cosmological constituents of the spacetime and here has the dimension of $\mathrm{m}^{-2}$. Accordingly, the lapse function \eqref{lapse} provides the exterior geometry of the CWBH. Note that, in the adopted geometric system of units, $Q$ and $\lambda$ have the dimensions of $\mathrm{m}$.} For $\lambda>Q$, this spacetime allows for two horizons; an event horizon $\mr_+$, and a cosmological horizon $\mr_{++}$ (instead of a Cauchy horizon), given by \cite{Fathi:2020b}
\begin{eqnarray}
&& \mr_+= \lambda \sin\left( {1\over 2} \arcsin\left(\frac{Q}{\lambda} \right)  \right),\label{w.6}\\
&& \mr_{++}=\lambda \cos\left( {1\over 2} \arcsin\left(\frac{Q}{\lambda}\right)\right).\label{w.7}
\end{eqnarray}
The extremal CWBH is obtained for $\lambda=Q$, whose causal structure is characterized by the unique horizon $\mathrm{r}_{\mathrm{ex}}=\mr_+=\mr_{++}=\lambda/\sqrt{2}$. The case of $\lambda<Q$ then corresponds to a naked singularity. {Note
that, to reduce the CWBH to the Reissner-Nordstr\"{o}m-(Anti-)de Sitter black hole of mass $M_0$, charge $Q_0$ and cosmological constant $\Lambda$, we need to let $\tilde{r}$ to be a free radial distance, $3\tilde{m}\rightarrow M_0$, $2\tilde{\varepsilon}\rightarrow\pm \Lambda$, and $Q\rightarrow 2 i Q_0$ (an imaginary transformation).}

\subsection{The rotating counterpart}

\begin{figure}[t]
    \centering
    \includegraphics[width=5.0cm]{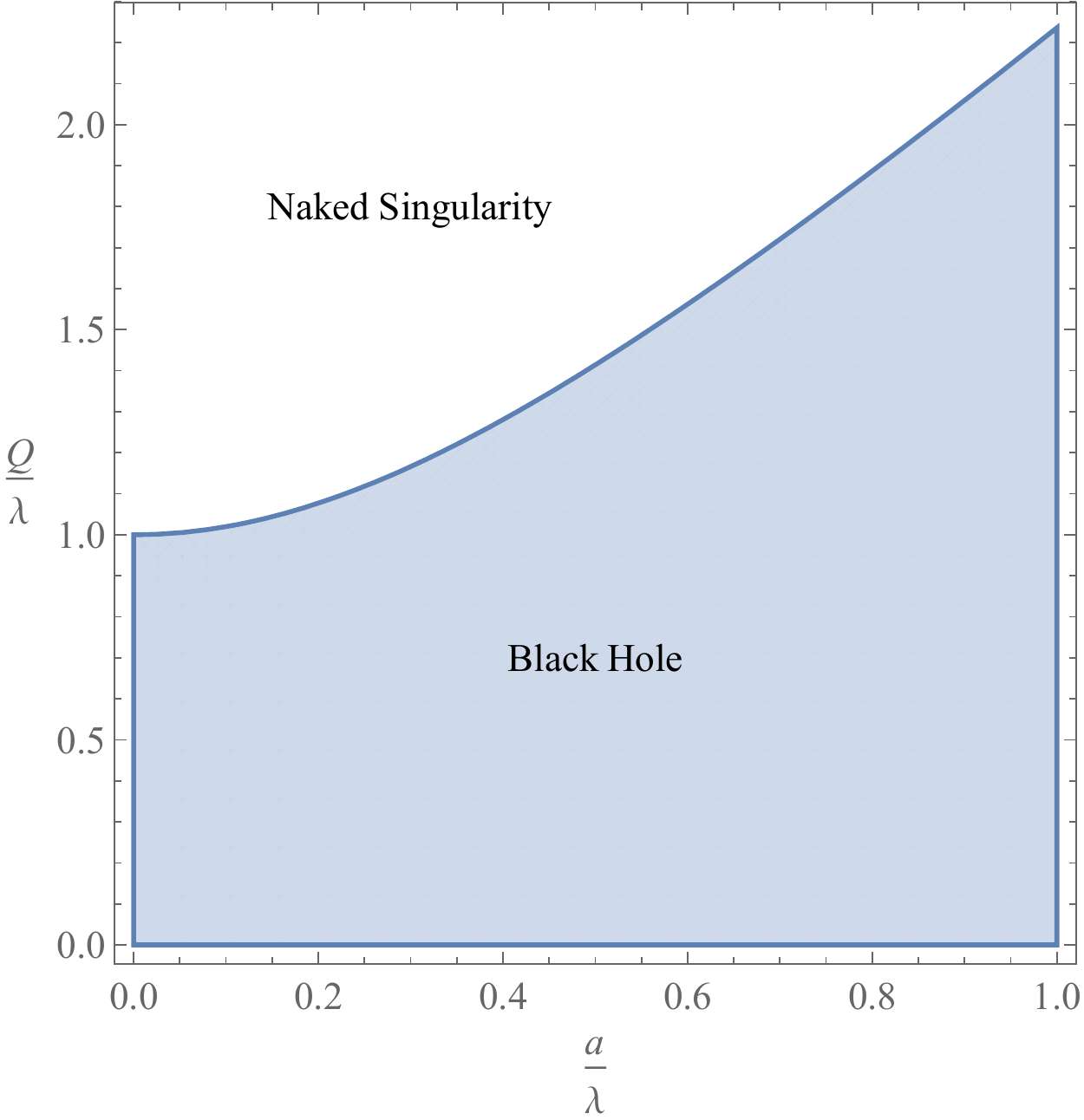}~(a)
    \includegraphics[width=5.0cm]{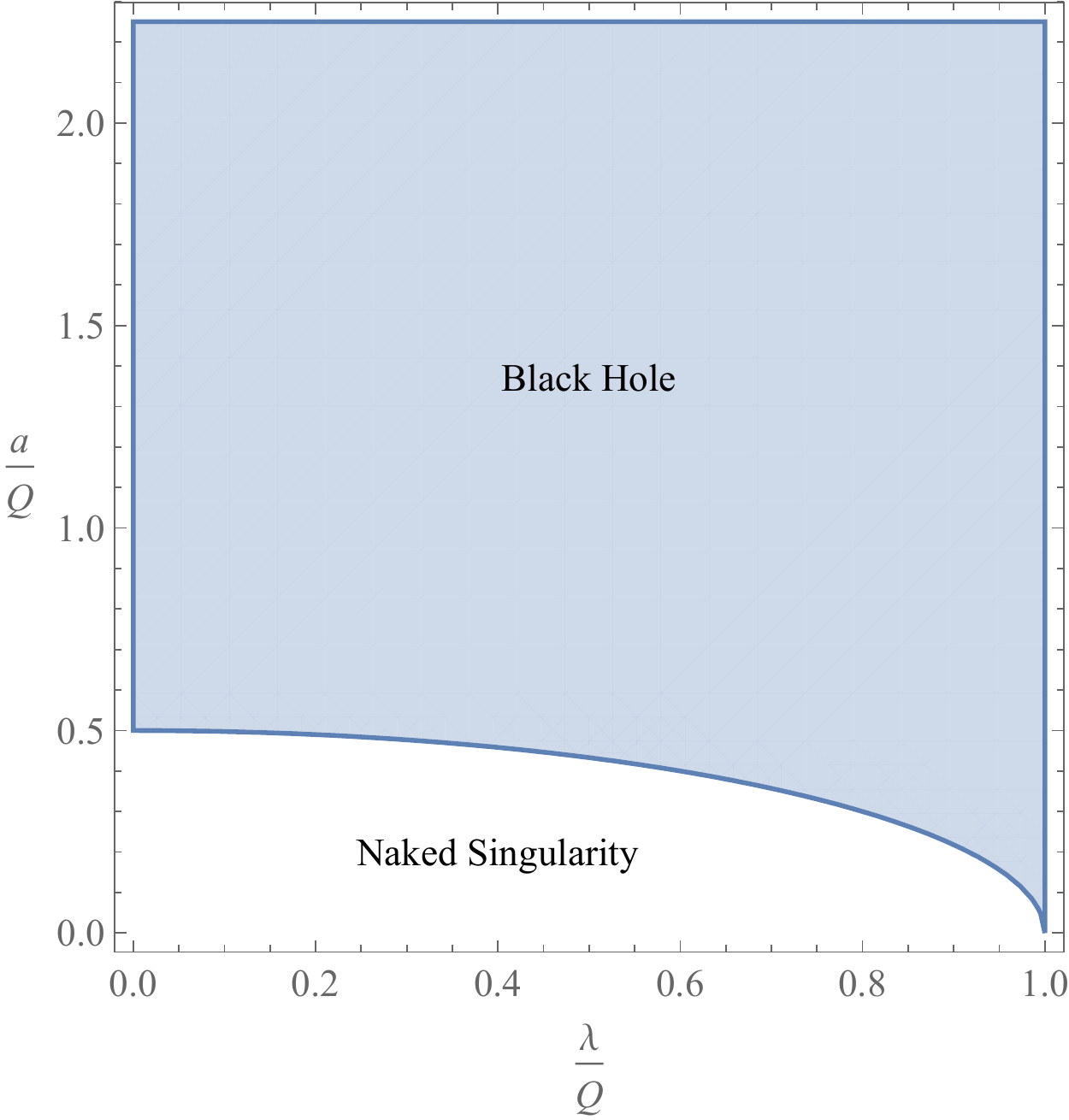}~(b)
    \includegraphics[width=5.0cm]{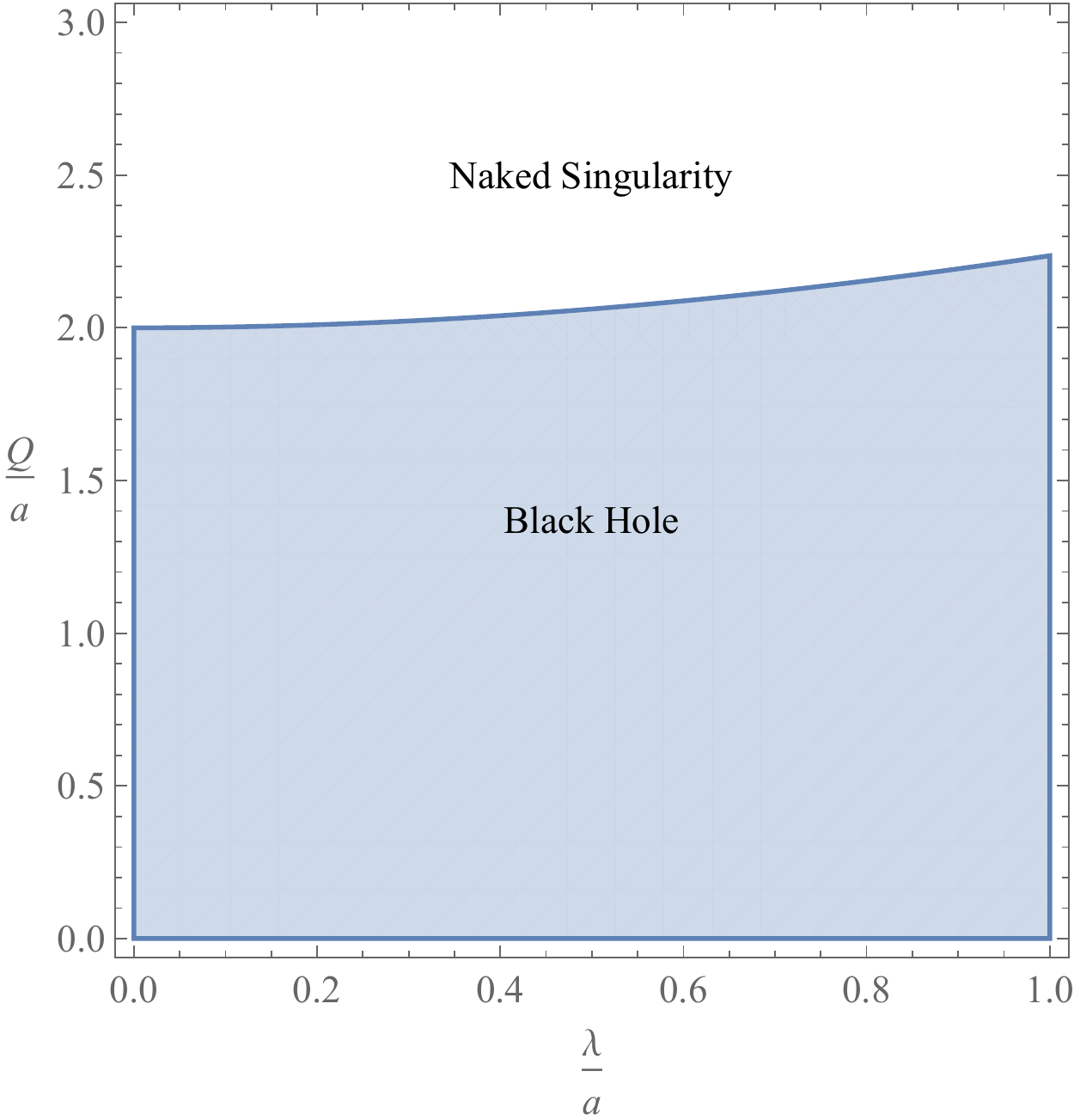}~(c)
    \caption{{The mutual sensitivity of the possibility of horizon formation to the pairs (a) $(Q,a)$, (b) $(a,\lambda)$, and (c) $(Q,\lambda)$. In the diagrams, the border between the regions of black hole and naked singularity, indicates the extremal black hole limit.} }
    \label{fig:BHRegion}
\end{figure}

\begin{figure}[t]
    \centering
    \includegraphics[width=5cm]{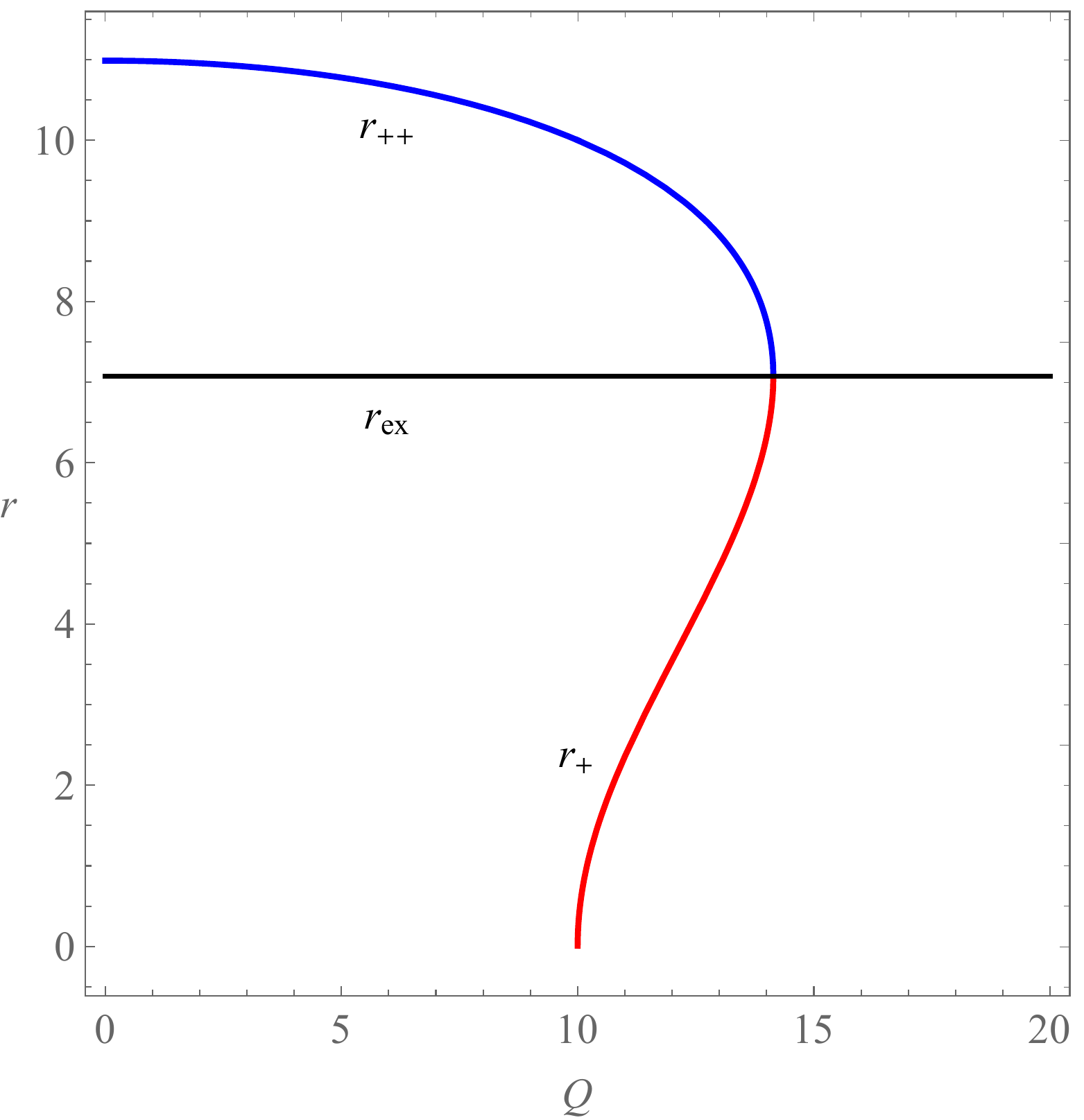}~(a)
    \includegraphics[width=5cm]{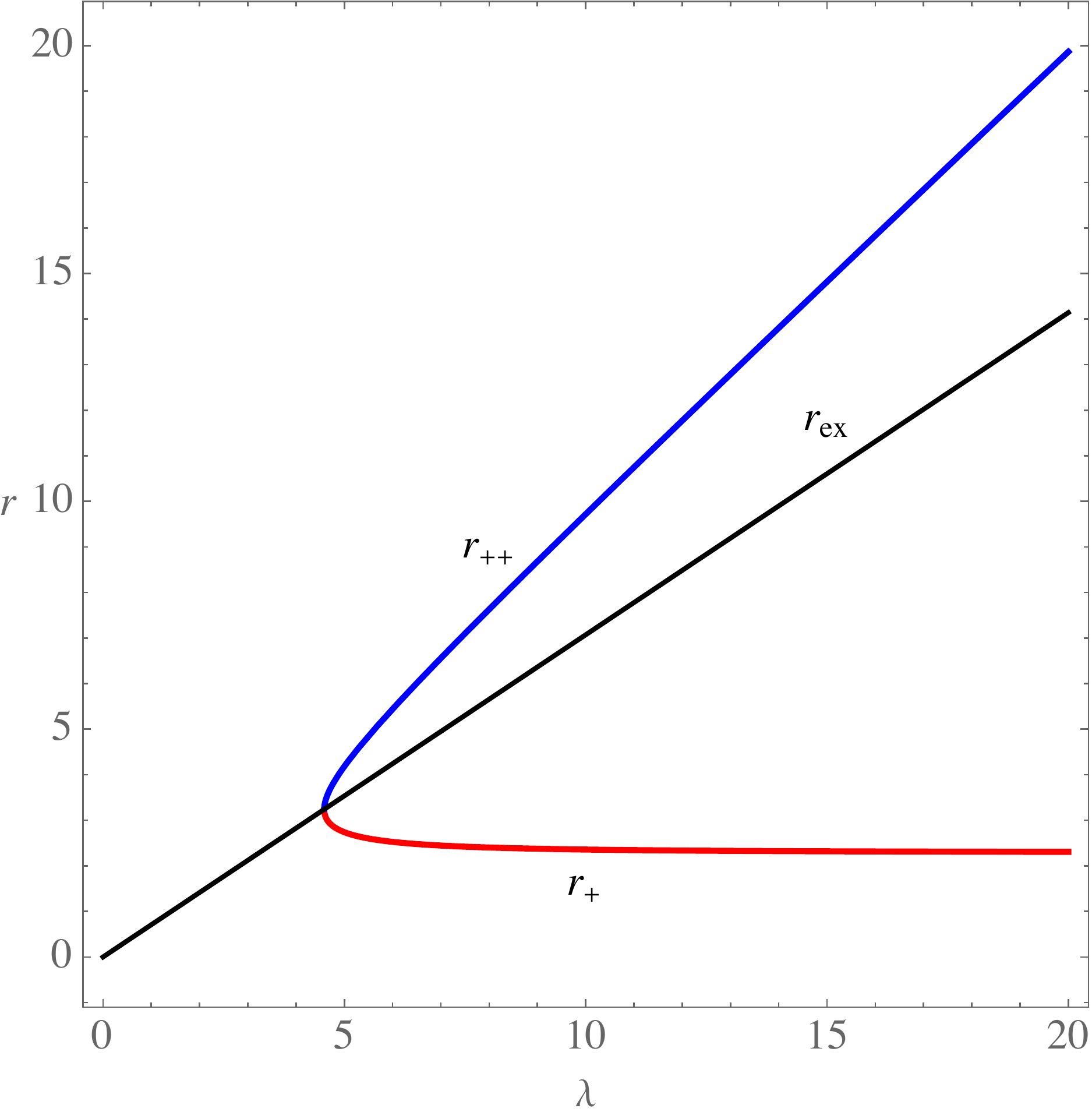}~(b)
    \includegraphics[width=5cm]{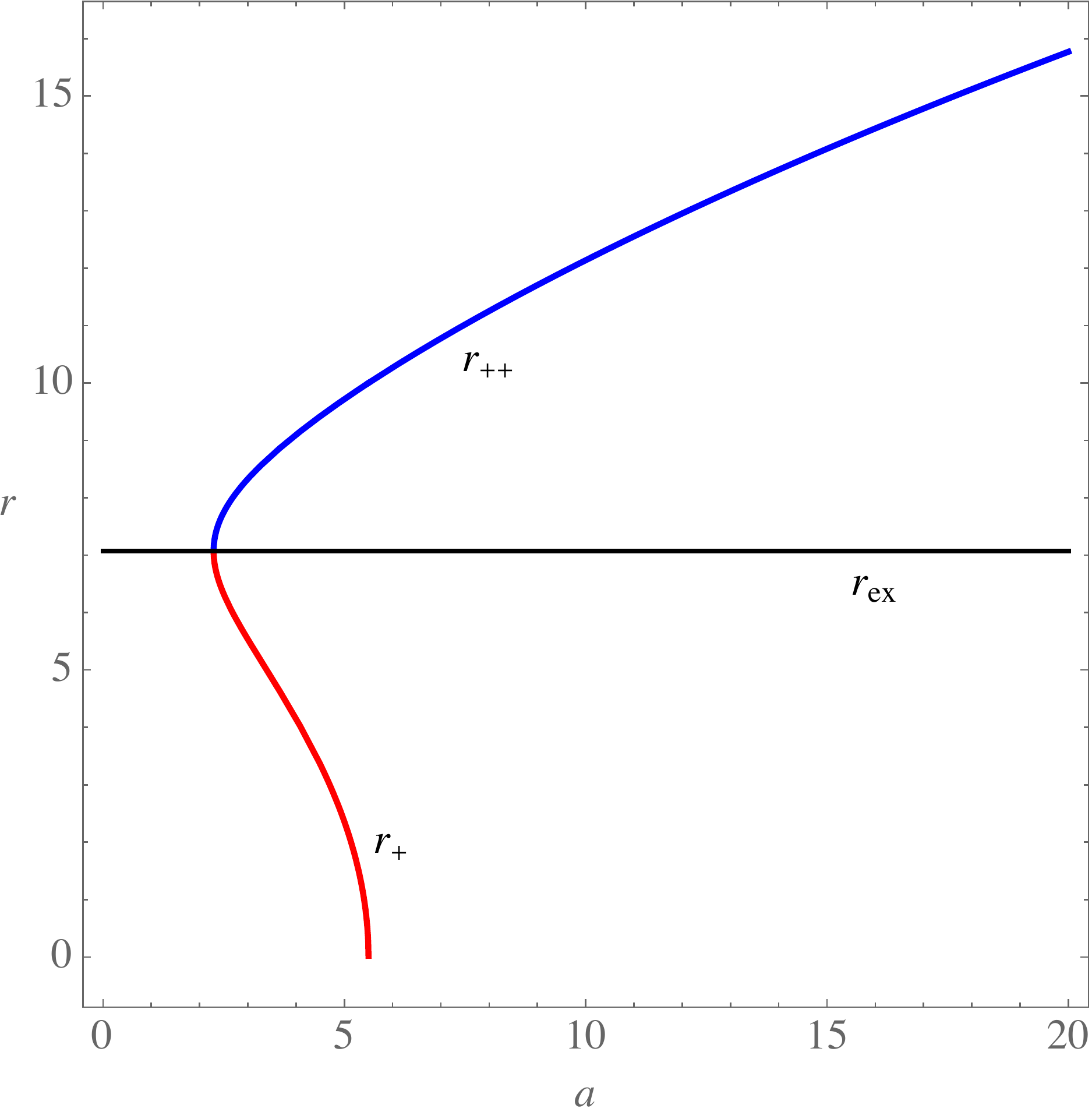}~(c)
    \caption{{The behavior of the horizons of the RCWBH, plotted for (a) $\lambda=10$ and $a=5$, (b) $Q=11$ and $a=5$, and (c) $\lambda=10$ and $Q=11$. } }
    \label{fig:rprpprex}
\end{figure}

To obtain the rotating counterpart of the CWBH, we tend to recast the CWBH solution in a stationary non-static form, in the Boyer-Lindquist coordinates. Such transformations are commonly done through the Newman-Janis algorithm of complexification of the radial coordinate \cite{NewmanJanis:1965}. This method has been widely used to generate stationary black hole solutions from their static counterparts\footnote{A more general version of the Newman-Janis method has been obtained by Shaikh in Ref.~\cite{Shaikh:2019}, which considers the seed static spacetimes of more generality.}. At first, this algorithm was intended to generate the Kerr solution from the Schwarzschild spacetime, and therefore, had been developed in the context of general relativity. It however, has had recent applications to black hole solutions in alternative theories of gravity \cite{Johannsen:2011_1,Bambi:2013_1,Moffat:2015_1,Jusufi:2020_1}. On the other hand, it has been pointed out that, when applied to alternative gravity, this method leads to distortions in the resultant rotating spacetimes, by introducing extra unknown sources in the original source \cite{Hansen:2013}. Hence, to obtain the rotating counterpart of the CWBH spacetime, we have applied the method proposed by Azreg-A\"{i}nou in Refs.~\cite{Azreg:2014,Azreg:2014_1}, which is the modified version of the Newman-Janis algorithm and has been applied successfully in producing imperfect rotating solutions from their static counterparts. {Employing this method for the line element \eqref{metr}, we obtain the following rotating spacetime: 
\begin{equation}\label{eq:metric_rotate}
    \ed s^2 = -\frac{\Xi}{\Sigma}\ed t^2 + \frac{\Sigma}{\Delta}\,\ed r^2+\Sigma\,\ed\theta^2-2 a \sin^2\theta\left(
    1-\frac{\Xi}{\Sigma}
    \right)\ed t\,\ed\phi
    + \sin^2\theta\left[
    \Sigma + a^2\sin^2\theta\left(
    2-\frac{\Xi}{\Sigma}
    \right)
    \right]\ed\phi^2,
\end{equation}
where
\begin{subequations}\label{eq:metr_comp}
\begin{align}
   & \Xi = \Delta-a^2\sin^2\theta,\\
   &\Delta = a^2 + r^2 B(r), \label{eq:Delta}\\
   &\Sigma = r^2+a^2\cos^2\theta,
\end{align}
\end{subequations}
in which, $a$ is the black hole's spin parameter, defined as $a = J/\tilde{m}$, with $J$ as the black hole's angular momentum. Therefore $a$ has the dimension of $\mathrm{m}$ in our geometric units. Also, the black hole's angular velocity is $\omega=-g_{t\phi}/g_{\phi\phi}$ \cite{Poisson:2009}.
{In fact, if the reference lapse function \eqref{eq:Referencef} is applied to Eq. \eqref{eq:Delta}, the components of the Bach tensor $W_{\alpha\beta}$, are vanished for the same expression of $f(r)$ as that in Eq. \eqref{eq:f(r)}. {We must however note that, this process can only be done by substitution of this lapse function in the components of the Bach tensor, applying a computer software }\footnote{{The calculation of the components of the Bach tensor has been done by the software Maple$^{{\small{\rm{TM}}}}$ 2018.}}. Hence, the line element \eqref{eq:metric_rotate} is a vacuum rotating solution to Weyl gravity, if a lapse function of the general form \eqref{eq:F_metric_0} is taken into account. Accordingly, for a massive charged spherical source, like the one assumed for the CWBH, the same constants can be obtained if the same method is pursued.} The only difference is that, the associated vector potential generated by the source, changes its form from that in Eq.~\eqref{eq:vectorPotential}, to \cite{Misner:1973}
\begin{equation}\label{eq:Arot}
    \tilde{A}_\alpha = \frac{q r}{\Sigma}\left(1,0,0, - a \sin^2\theta\right).
\end{equation}
The exterior geometry of the RCWBH is therefore specified by applying the lapse function \eqref{lapse}, that provides
\begin{equation}\label{eq:DeltaRCWBH}
    \Delta = a^2 +r^2-\frac{r^4}{\lambda^2}-\frac{Q^2}{4}.
\end{equation}}
As for the spherically symmetric stationary spacetimes, the RCWBH admits two Killing vectors $\bm{\xi}_{(t)}$ and $\bm{\xi}_{(\phi)}$, satisfying
\begin{eqnarray}\label{eq:Killings_0}
&&\xi^\alpha_{(t)}\xi_{\alpha(t)} = g_{tt},\label{eq:Killings_0_t}\\
&&\xi^\alpha_{(\phi)}\xi_{\alpha(\phi)} = g_{\phi\phi}\label{eq:Killings_0_phi},
\end{eqnarray}
that correspond to the translational and rotational symmetries, and the relevant invariants of motion. The black hole's event and cosmological horizons, are now obtained by solving $g^{rr} = 0$, which results in (see appendix \ref{app:Ap})
\begin{eqnarray}\label{eq:horizons_new}
   && r_+ = \lambda \sin\left(\frac{1}{2}\arcsin\left(\frac{2}{\lambda}\sqrt{\frac{Q^2}{4}-a^2}\right)\right),\label{eq:horizons_new_rp}\\
  && r_{++} = \lambda\cos\left(\frac{1}{2}\arcsin\left(\frac{2}{\lambda}\sqrt{\frac{Q^2}{4}-a^2}\right)\right)\label{eq:horizons_new_rpp},
\end{eqnarray}
for which, the extremal black hole horizon $r_\mathrm{ex} = \lambda/\sqrt{2}$ is obtained for $Q_{\mathrm{ex}} = \pm\sqrt{4 a^2+\lambda ^2}$, and a naked singularity appears for $Q>\sqrt{4a^2+\lambda^2}$ (for $Q>0$). As shown in Fig.~\ref{fig:BHRegion}, not all values of $Q$, $\lambda$ and $a$ are allowed to construct a black hole (censored region). Also, as shown in Fig.~\ref{fig:rprpprex}, for fixed $\lambda$ and $a$, increase in $Q$ increases the size of $r_+$ and decreases that of $r_{++}$, until they coincide on $r_{+}=r_{++}=r_{\mathrm{ex}}$ at $Q=Q_{\mathrm{ex}}$. {Same happens for decrease in $\lambda$ for fixed $Q$ and $a$, that leads to $\lambda=\lambda_\mathrm{ex}$, and decrease in $a$ for fixed $\lambda$ and $Q$, leading to $a=a_\mathrm{ex}$}. It is also easy to show that the hypersurfaces corresponding to $r_+$ and $r_{++}$, are null. We continue analyzing the RCWBH in the next section, regarding the stationary and static observers.

\section{The Ergosphere}\label{sec:ergosphere}

\begin{figure}[t]
    \centering
    \includegraphics[width=7cm]{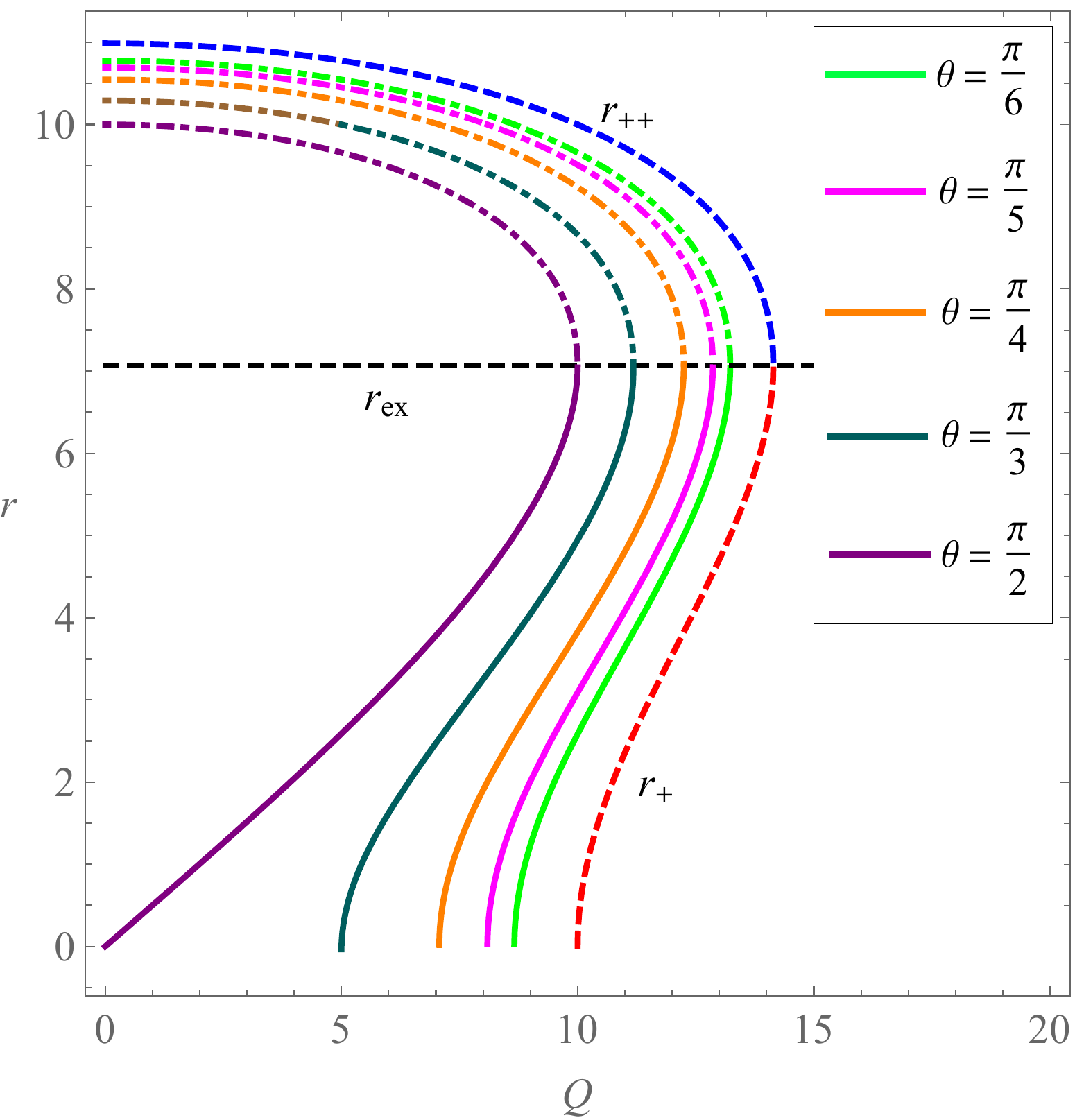}
    \caption{The behavior of $r_{SL}$, with respect to changes in $Q$. The plots have been done for $\lambda=10$ and $a=5$ and five different values of $\theta$. The solid and dot-dashed curves correspond respectively to $r_{SL1}$ and $r_{SL2}$, and the black hole horizons have been shown with dashed lines.}
    \label{fig:rSL}
\end{figure}

\begin{figure}[t]
    \centering
    \includegraphics[width=5.4cm]{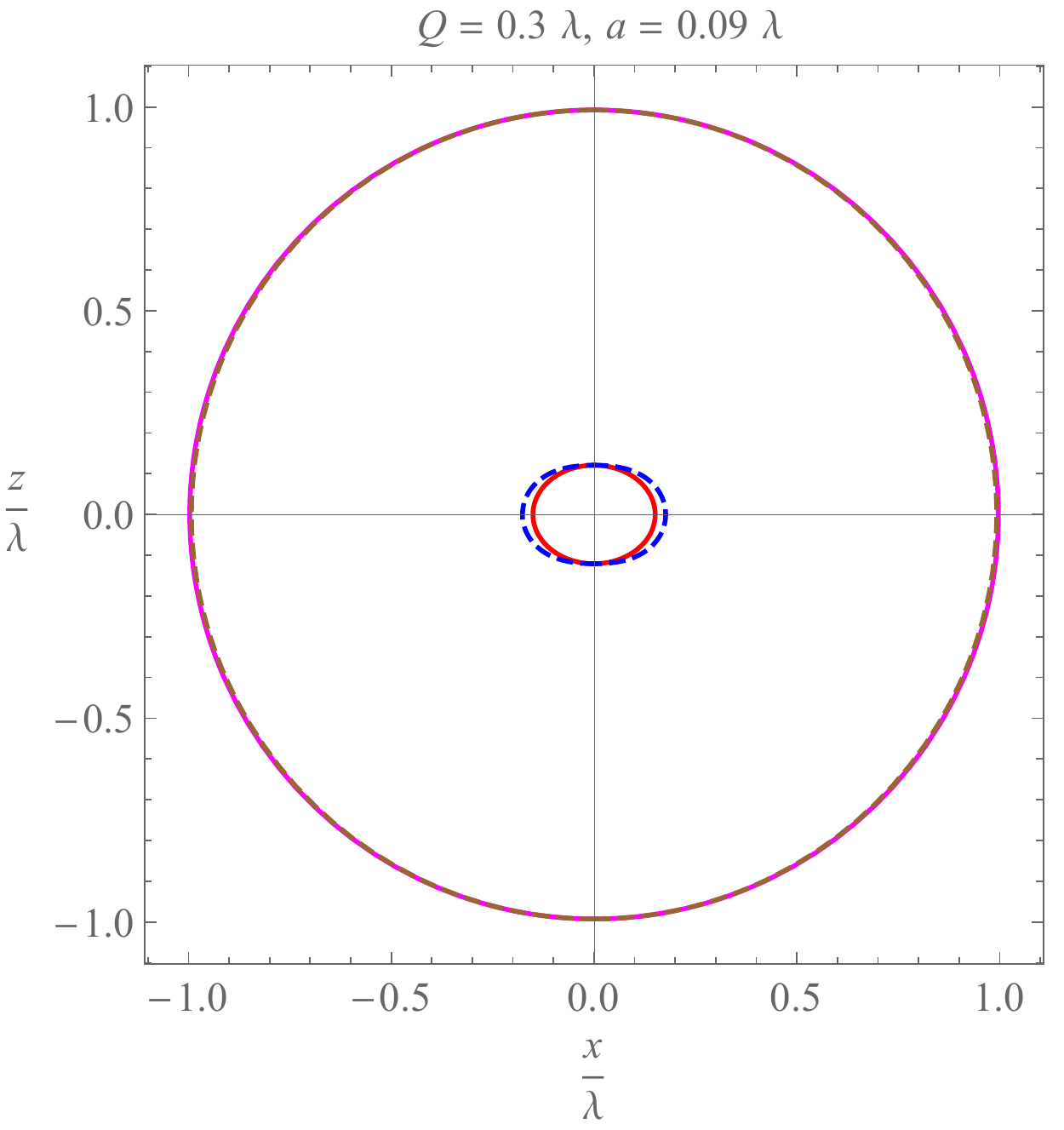} (a)
    \includegraphics[width=5.4cm]{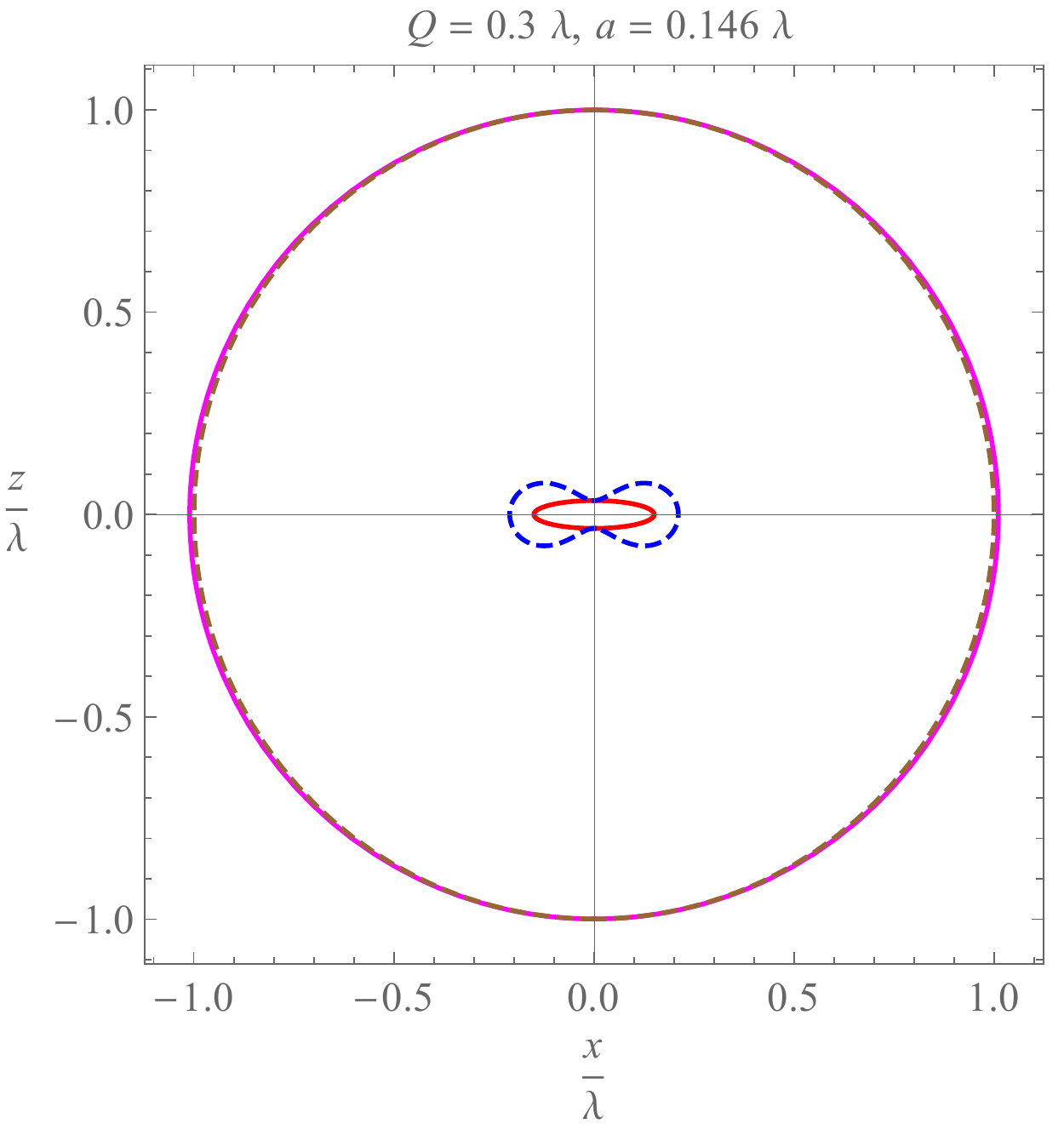} (b)
    \includegraphics[width=5.4cm]{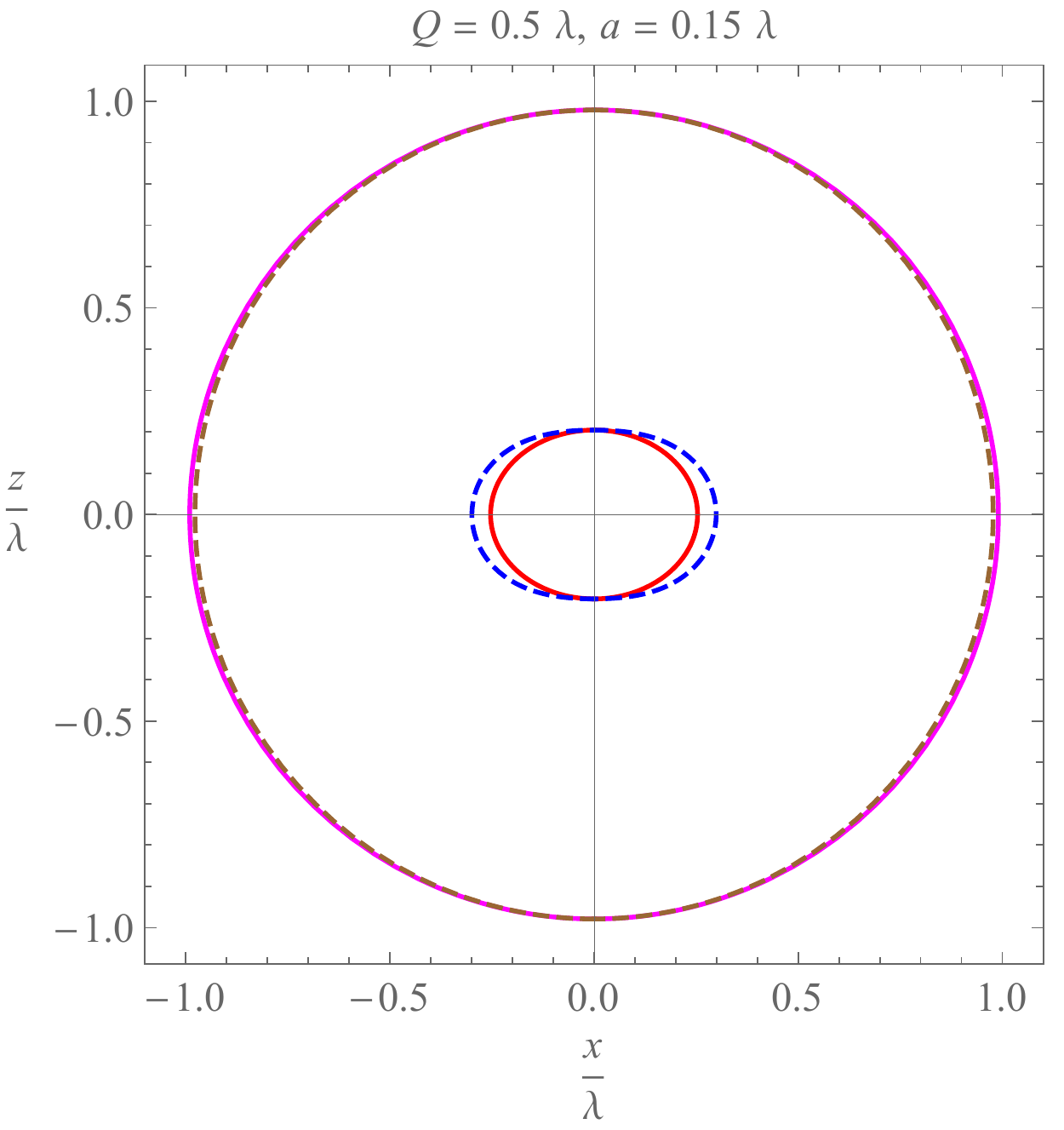} (c)
    \includegraphics[width=5.4cm]{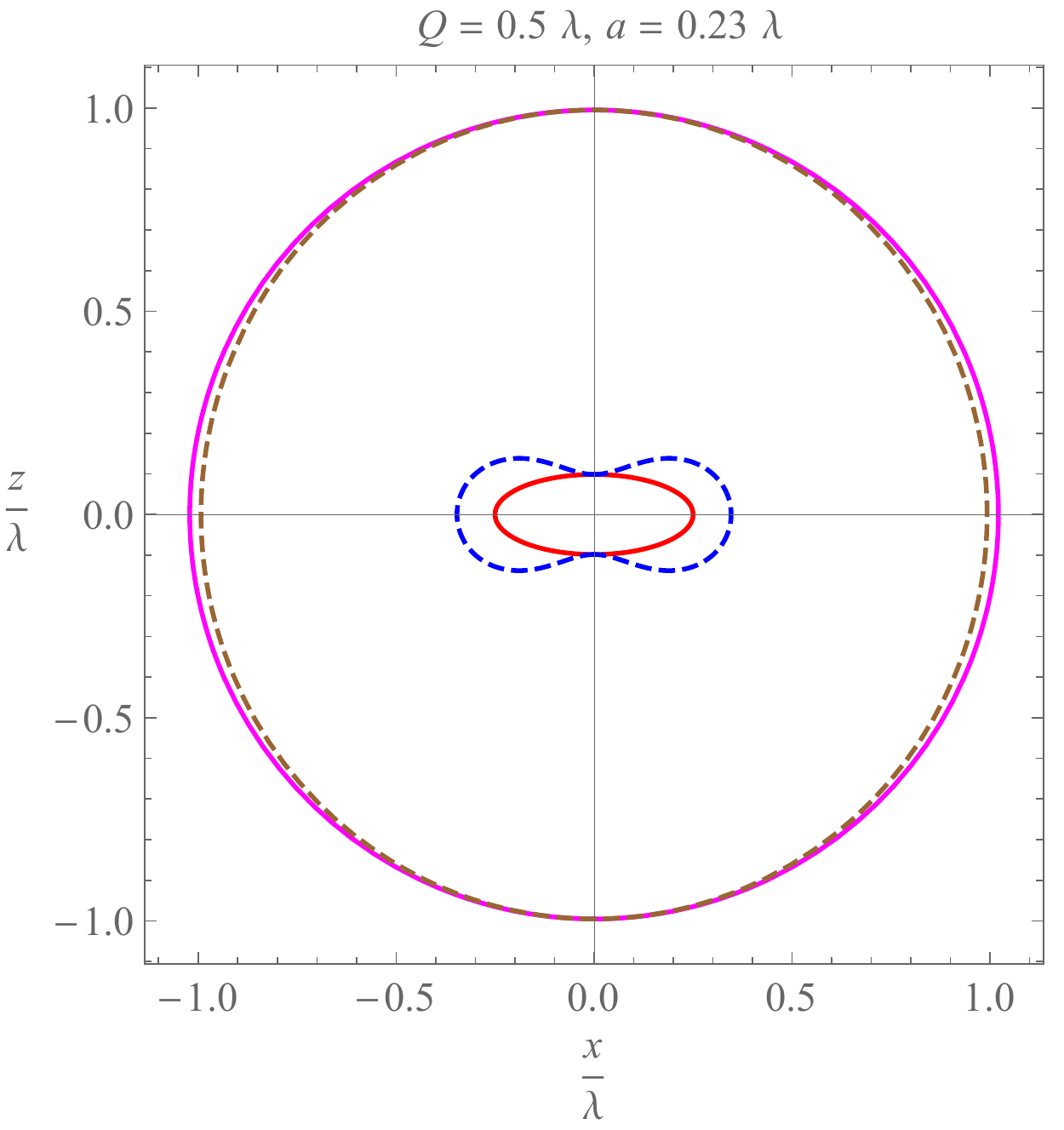} (d)
    \includegraphics[width=5.4cm]{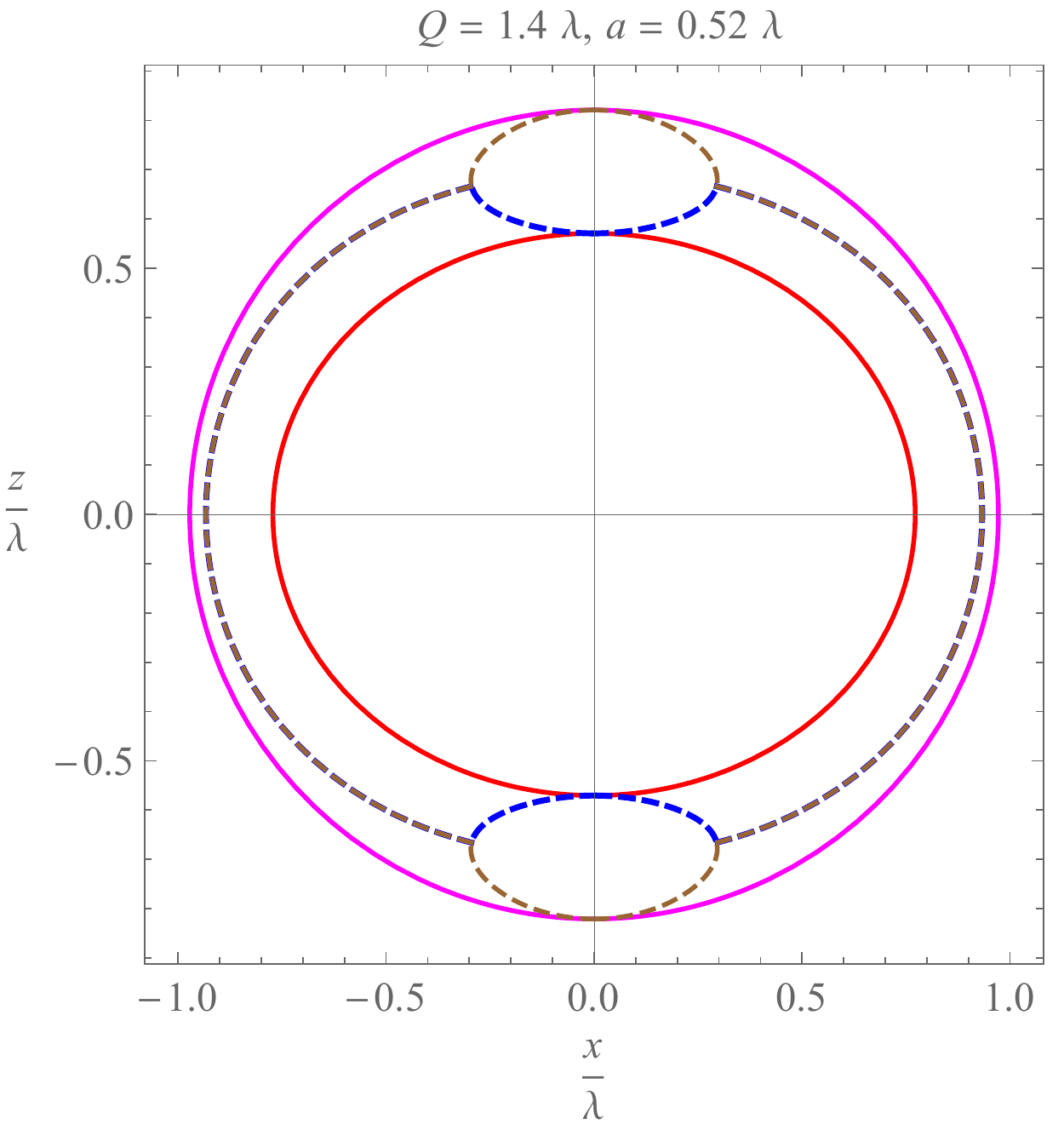} (e)
    \includegraphics[width=5.4cm]{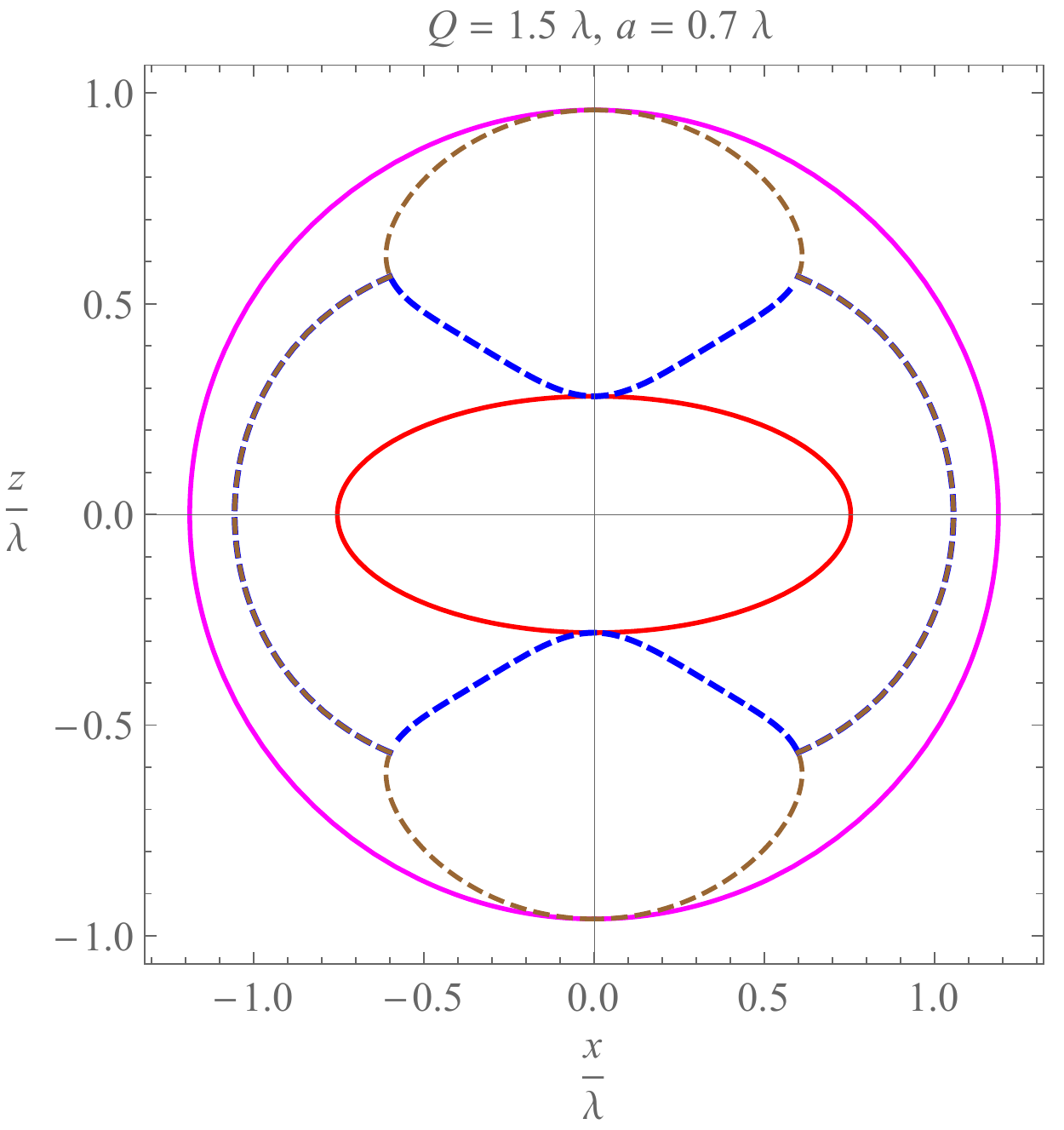} (f)
    \includegraphics[width=5.4cm]{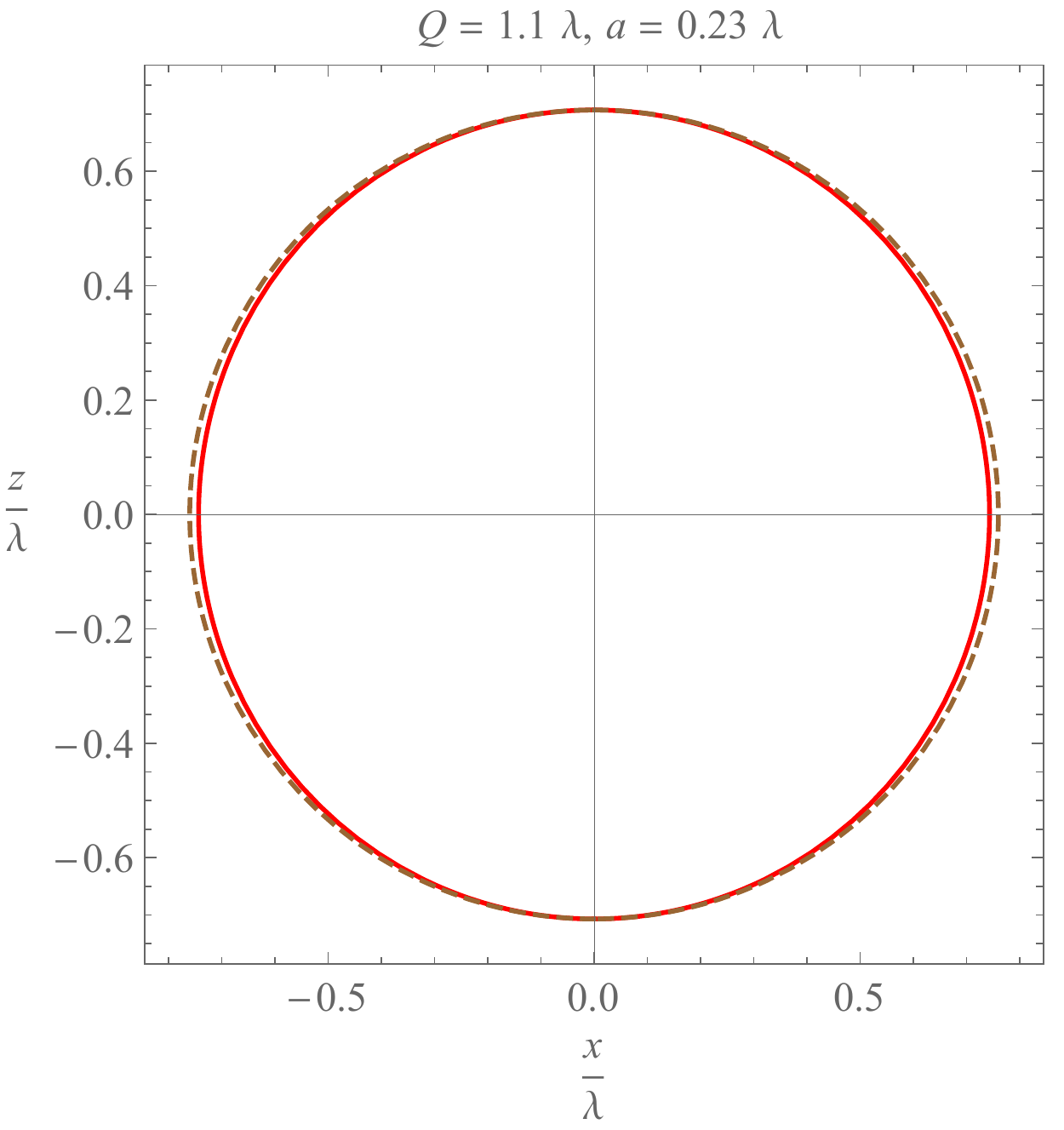} (g)
    \includegraphics[width=5.4cm]{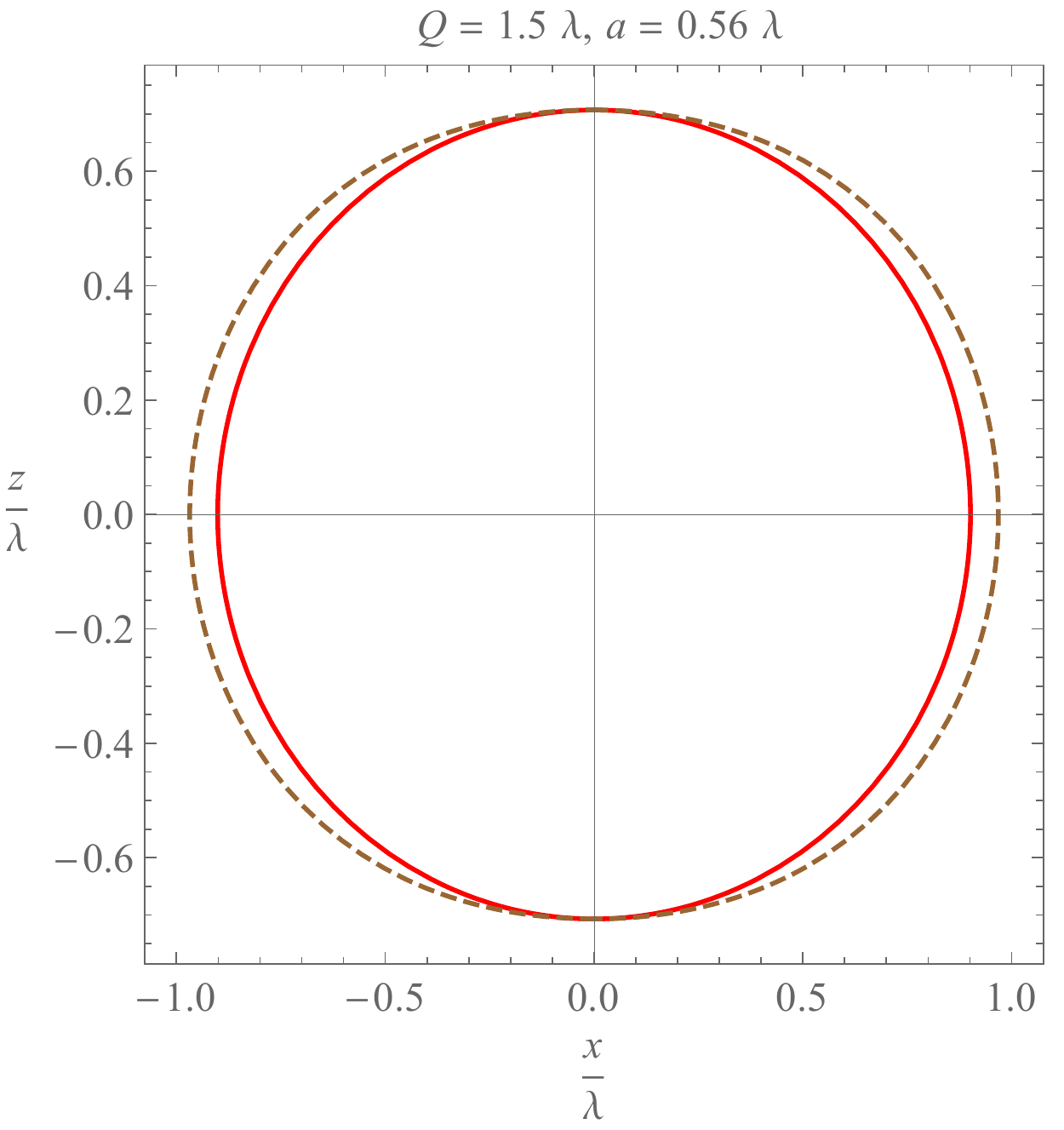} (h)
    \includegraphics[width=5.4cm]{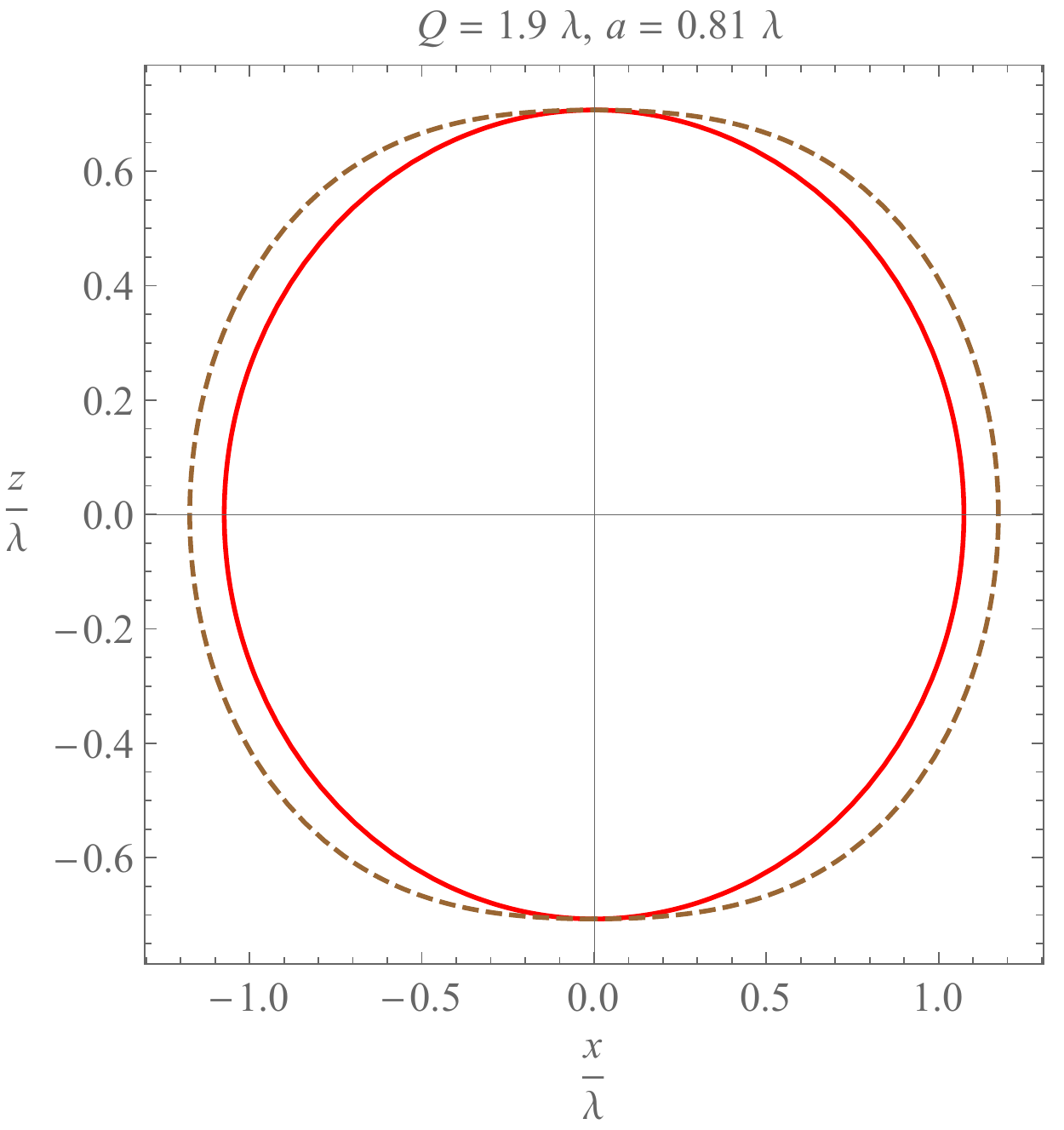} (i)
    \caption{The hypersurfaces corresponding to the horizons and the static limit, plotted for $\lambda=10$ and several fixed values of $Q$ and $a$, as viewed from the $y$ axis. The axes have been given in terms of dimension-less values $z/\lambda$ and $x/\lambda$. In the (a)-(f) diagrams, the smaller and the larger solid contours correspond respectively to $r_+$ and $r_{++}$, whereas the smaller and the larger dashed ones relate to $r_{SL1}$ (the static limit) and $r_{SL2}$. The region $r_+<r<r_{SL1}$ indicates the ergosphere in each of the cases. The figures have been scaled in a way that the $r_{++}$ surface appears as a complete circle. The (g)-(i) diagrams demonstrate the ergoregion in the case of the extremal black hole. The solid and the dashed contours correspond respectively to $r_\mathrm{ex}$ and $r_{SL(\mathrm{ex})}$.}
    \label{fig:ergosphere}
\end{figure}

Beside being encompassed by the event horizon, the rotating black holes are also characterized by another hypersurface, which is formed as a result of the black hole's spin and limits the existence of static observers. This surface is therefore identified with the event horizon, only in the limit of the vanishing spin parameter \cite{Bardeen:1972a,Bardeen:1973a,Chandrasekhar:579245}. In this section, we determine this surface and discuss it analytically as well as illustratively. However, let us firstly consider an important feature of rotating spacetimes, by considering the zero angular momentum observers (ZAMOs), with the vanishing angular momentum defined as \cite{Poisson:2009}
\begin{eqnarray}\label{eq:ZAMOsL}
    \tilde{L} &\equiv& g_{\phi\alpha}\frac{\ed x}{\ed\tau}^\alpha \nonumber\\
     &=& g_{\phi t}\frac{\ed t}{\ed\tau}  + g_{\phi\phi}\frac{\ed\phi}{\ed\tau} = 0,
\end{eqnarray}
in which, $\tau$ is the proper time associated with the observers (or particles) and throughout this paper, we assign it to the trajectory parameter. Accordingly, the observer's angular velocity is obtained from Eqs.~\eqref{eq:metric_rotate} and \eqref{eq:metr_comp}, reading as
\begin{equation}\label{eq:ZAMOsOmega}
    \Omega = \frac{\ed\phi}{\ed t} = -\frac{g_{t\phi}}{g_{\phi\phi}} = \frac{a\left(r^2+a^2-\Delta\right)}{\left(r^2+a^2\right)^2-\Delta a^2\sin^2\theta},
\end{equation}
which is the same as the angular velocity of the rotating black hole and increases as $r$ decreases, until it reaches its maximum value
\begin{equation}\label{eq:Omega_max}
    \Omega_{\max}=\Omega|_{r=r_+}\equiv\Omega_H= \frac{a}{r_+^2+a^2} = \omega_+,
\end{equation}
where $\omega_+ \equiv \omega(r_+)$ is the black hole's angular velocity at its event horizon. For a slowly rotating black hole (small $a$), $\Omega_H$ reduces to
\begin{equation}\label{eq:Omega_max_slow}
 \Omega_H \approx \frac{a}{\lambda^2\sin^2\left(
 \frac{1}{2}\arcsin\left(
 \frac{Q}{\lambda}
 \right)
 \right)}.
\end{equation}
Therefore, on the event horizon, ZAMOs move in the same direction as the black hole’s own rotation (also called \textit{corotation}), as a result of dragging of their inertial frames. Aside from the ZAMOs, now to deal with the static observers, let us consider their velocity four-vector as
\begin{equation}\label{eq:u_static}
    u^\alpha = \mathfrak{n}\, \xi^\alpha_{(t)},
\end{equation}
with the normalization factor $\mathfrak{n} = (-g_{\alpha\beta}\xi^\alpha_{(t)}\xi^\beta_{(t)})^{-1/2}=1/\sqrt{-g_{tt}}$, according to Eq.~\eqref{eq:Killings_0_t}. Such observers cannot exist everywhere in the spacetime, and they are indeed confined to a limit defined in terms of the validity of Eq.~\eqref{eq:u_static}. In fact, for $\mathfrak{n}^{-2}=-g_{tt}=0$, this equation breaks down and $\bm{\xi}_{(t)}$ becomes null. This way, a \textit{static limit}\footnote{The surface corresponding to the static limit, is also called the surface of infinite redshift \cite{Ryder:2009}.} is obtained, whose radius, $r_{SL}$, is calculated by solving the equation $g_{tt}=0$. This equation yields the two positive solutions
\begin{eqnarray}\label{eq:rSL}
  &&  r_{SL1} = \lambda \sin\left(\frac{1}{2}
  \arcsin\left(
  \frac{2}{\lambda}\sqrt{\frac{Q^2}{4}-a^2\cos^2\theta}
  \right)
  \right),\label{eq:rSL1}\\
  &&  r_{SL2} = \lambda \cos\left(\frac{1}{2}
  \arcsin\left(
  \frac{2}{\lambda}\sqrt{\frac{Q^2}{4}-a^2\cos^2\theta}
  \right)
  \right),\label{eq:rSL2}
\end{eqnarray}
that satisfy the condition of causality $r_+<r_{SL1}<r_{SL2}<r_{++}$. The domain $r<r_{SL1}$, corresponds to a region, in which, the static observers can no longer remain static\footnote{In other words, they will no longer exist.}, and the so-called \textit{frame-dragging} effect forces them to rotate with the black hole. In Fig.~\ref{fig:rSL}, the behavior of the above two solutions has been plotted for a definite value of $a$. 
As the angular parameter increases, the static limit surface (corresponding to $r_{SL1}$) recedes from the event horizon, and hence, the ergosphere expands. On the other hand, $r_{SL2}$ shrinks under the same conditions. Note that, in the case of an extremal black hole ($Q^2=4 a^2 + \lambda^2$), the above surfaces coincide at
\begin{equation}\label{eq:SLext}
   r_{SL(\mathrm{ex})} = \lambda  \cos \left(\frac{\pi}{4}-\frac{1}{4} 
    \arccos\left(\frac{8 a^2\sin^2\theta+\lambda ^2}{\lambda ^2}\right)\right).
\end{equation}
Comparing Eqs. \eqref{eq:horizons_new_rp} and \eqref{eq:rSL1}, we can notice that for $\theta\neq n \pi~ (n=0,\pm1,\pm2,...)$,  the surface of static limit does not coincide with that of the event horizon, and these surfaces, together, form an \textit{ergosphere} (or \textit{ergoregion}). To demonstrate the ergosphere of the black hole, let us introduce the Cartesian coordinates \cite{Boyer:1967} 
\begin{subequations}\label{eq:Cartesian}
\begin{align}
  &  x = \sqrt{r^2+a^2}\sin\theta\cos\phi,\\
  &  y = \sqrt{r^2+a^2}\sin\theta\sin\phi,\\
  &  z = r\cos\theta,
\end{align}
\end{subequations}
which are defined in terms of the Boyer-Lindquist coordinates of the line element \eqref{eq:metric_rotate}. Accordingly, the horizon hypersurfaces can be plotted, as being observed either from the coordinate of axial symmetry (i.e. in the $x$-$y$ plane), or from the $y$ (or $x$) coordinate (i.e. in the $z$-$x$ (or $z$-$y$) plane). In Fig.~\ref{fig:ergosphere}, the horizons and the static limit surfaces have been plotted for several values of $a$ and $Q$ within the allowed values of Fig.~\ref{fig:BHRegion}, and therefore, the ergosphere for each of these cases has been demonstrated. 
{As it is seen in the figures, increase in $a$ for each fixed value of $Q$, stretches the event horizon's cross-section and acts in favor of dividing the ergosphere into separate regions at each side of the event horizon. In the special cases of $Q>\lambda$, $r_{SL1}$ and $r_{SL2}$ match in certain regions outside the event horizon and form ergospheres of peculiar shapes. For the extremal black hole, it is naturally $Q>\lambda$ and the spacetime encounters one horizon and one static limit, and its ergosphere increases in size as the distance between the values of $Q$ and $a$ increases.} In fact, stationary observers of angular velocity $\Omega$ and the four-velocity $u_s^\alpha = \mathfrak{n}_s \xi_s^\alpha$, with
\begin{eqnarray}\label{eq:KillingStationary}
 && \xi_s^\alpha = \xi_{(t)}^\alpha + \Omega\xi_{(\phi)}^a,\\
 &&  \mathfrak{n}_s^{-2} = -g_{\alpha\beta}\xi_s^\alpha \xi_s^\beta = -g_{\phi\phi}\left(
 \Omega^2-2\omega\Omega+\frac{g_{tt}}{g_{\phi\phi}}
 \right),
\end{eqnarray}
can exist in the ergosphere, as long as the Killing vector $\bm{\xi}_s$ remains time-like. This corresponds to $\mathfrak{n}_s^{-2}>0$, or $\Omega_-<\Omega<\Omega_+$, with
\begin{equation}\label{eq:Omegapm}
    \Omega_\pm = \omega\pm\frac{\sqrt{\Delta}\sin\theta}{g_{\phi\phi}}.
\end{equation}
Accordingly, this vector becomes null on the event horizon\footnote{This also proves that the event horizon is a Killing horizon.}. Therefore, by approaching the event horizon, $\Omega\rightarrow\Omega_H\equiv\Omega_+$, and the particles will be in the state of corotation with the black hole. Such particles encounter the surface gravity \cite{Poisson:2009}
\begin{equation}\label{eq:surface_gravity_formular}
    \kappa = \left.\frac{\Delta'(r)}{2(r^2+a^2)}\right|_{r=r_+},
\end{equation}
at the vicinity of the event horizon. Applying Eqs.~\eqref{eq:Delta} and  \eqref{eq:horizons_new_rp}, and after a little algebra, this gives
\begin{equation}\label{eq:surface_gravty_rp}
\kappa = \frac{r_+}{a^2+r_+^2}\sqrt{\lambda^2-Q^2+4a^2}.
\end{equation}
For the extremal black hole ($r_+\rightarrow r_\mathrm{ex}$), we have from Eq.~\eqref{eq:surface_gravity_formular} that $\kappa_\mathrm{ex}=0$. Note that, for asymptotically flat spacetimes, where the energy can be defined relative to an observer located at infinity, the ergosphere is highlighted also by a theoretical scenario, termed as the \textit{Penrose process}, through which, particles of {negative-energy} created in the ergosphere can extract positive rotational energy from a rotating black hole \cite{Penrose:2002}.\\

Note that, the optical appearance of a black hole to distant observers, does not rely on the positioning of its ergosphere, since this latter, affects only the time-like particles. The conceived image of a black hole is, in fact, a shadow that is confined by particular photons on unstable (critical) orbits. Such photons construct a photon surface around the black hole. For the case of the RCWBH, this will be given a detailed discussion in the next section.

\section{Shadow of the black hole}\label{sec:shadow}

The light propagation around black holes is of remarkable  importance in astrophysics, specially because of the evidences that can be achieved by advancements in the field of observational astronomy. Photons that lie on unstable orbits in the gravitational field of the black holes, will either fall onto the event horizon or escape to infinity. To the observer, these latter ones constitute a bright photon ring which confines the black hole shadow \cite{Synge:1966,Cunningham:1972,Bardeen:1973a,Luminet:1979}. In particular, the Luminet's optical simulation of a Schwarzschild black hole and its accretion in 1979 \cite{Luminet:1979}, gave more insights about the photon rings, resulting from the extremely warped region around the black holes. The formulations obtained in this way, then helped scientists to confine the shadow of rotating black holes inside their respected photon rings. Accordingly, the mathematical methods to calculate the form and size of a Kerr black hole's shadow were then developed by Bardeen \cite{Bardeen:1972a,Bardeen:1973a,Bardeen:1973b}, and the same methods were reused by Chandrasekhar  \cite{Chandrasekhar:579245}. These methods were later developed and generalized widely (see Refs.~\cite{Bray:1986,Vazquez:2004,Grenzebach:2014,Grenzebach:2016,Perlick:2018a,Kogan:2018}). Having these methods in hand, a large number of black hole spacetimes, including those with cosmological components, were given rigorous analytical calculations, simulations, numerical, and observational studies \cite{Vries:1999,Shen:2005,Amarilla:2010,Amarilla:2012,Yumoto:2012,Amarilla:2013,Atamurotov:2013,Abdujabbarov:2015,Abdujabbarov:2016,Amir:2018,Tsukamoto:2018,Cunha:2018,Mizuno:2018,Mishra:2019,Kumar:2020b}. The black hole shadow is of great importance as it provides information about the light propagation in near-horizon regions. Recently, several investigations have been devoted to establish relations between the shadow and black hole parameters (see for example Refs.~\cite{Zhang:2020,Belhaj:2020} for thermodynamic, and Refs.~\cite{Kramer:2004,Psaltis:2008,Harko:2009,Psaltis:2015,Johannsen:2016a,Psaltis:2019,Dymnikova:2019,Kumar:2020a} for dynamical aspects of rotating black holes). 

\begin{figure}[t]
    \centering
    \includegraphics[width=5.4cm]{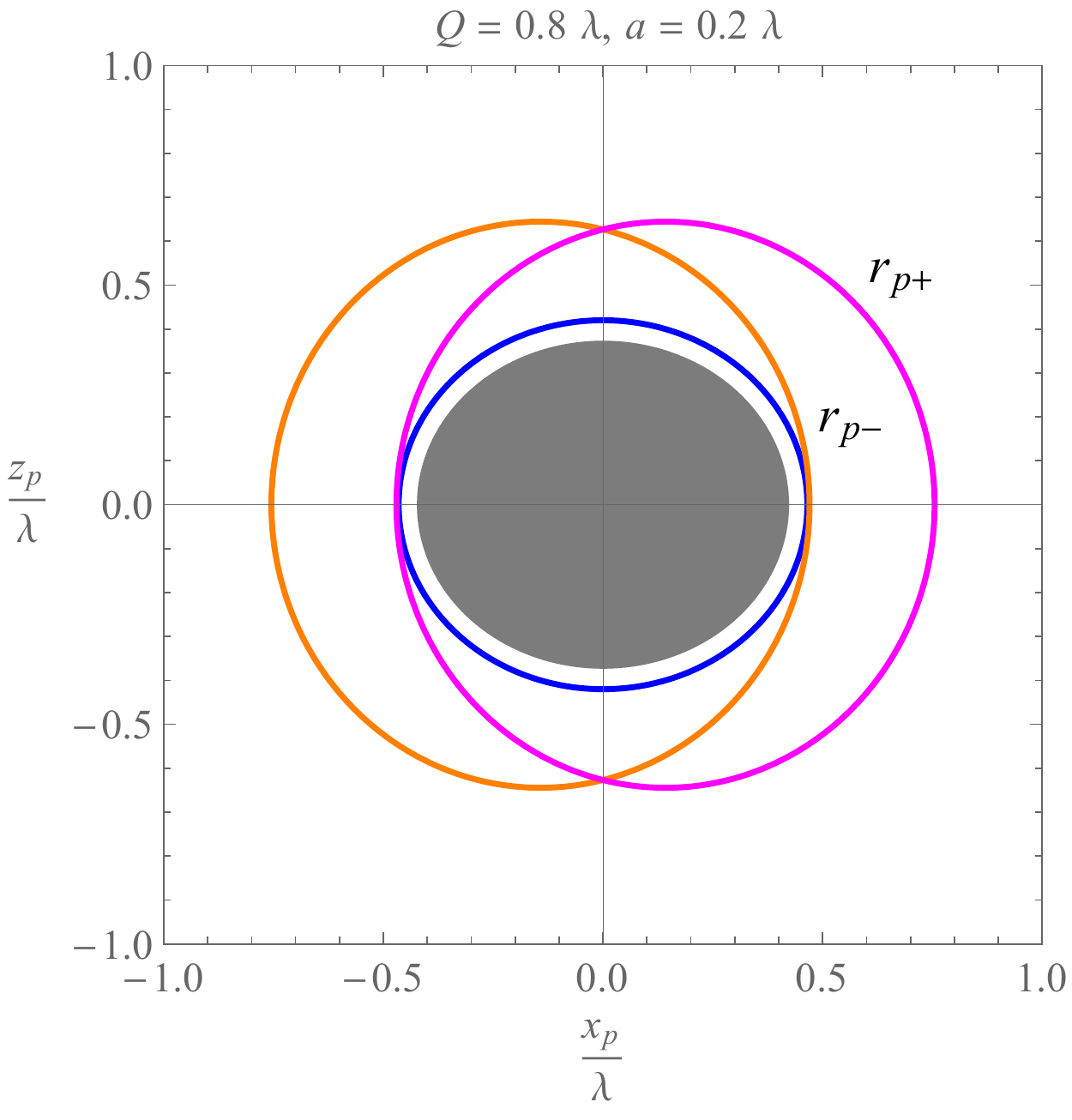}~(a)
    \includegraphics[width=5.4cm]{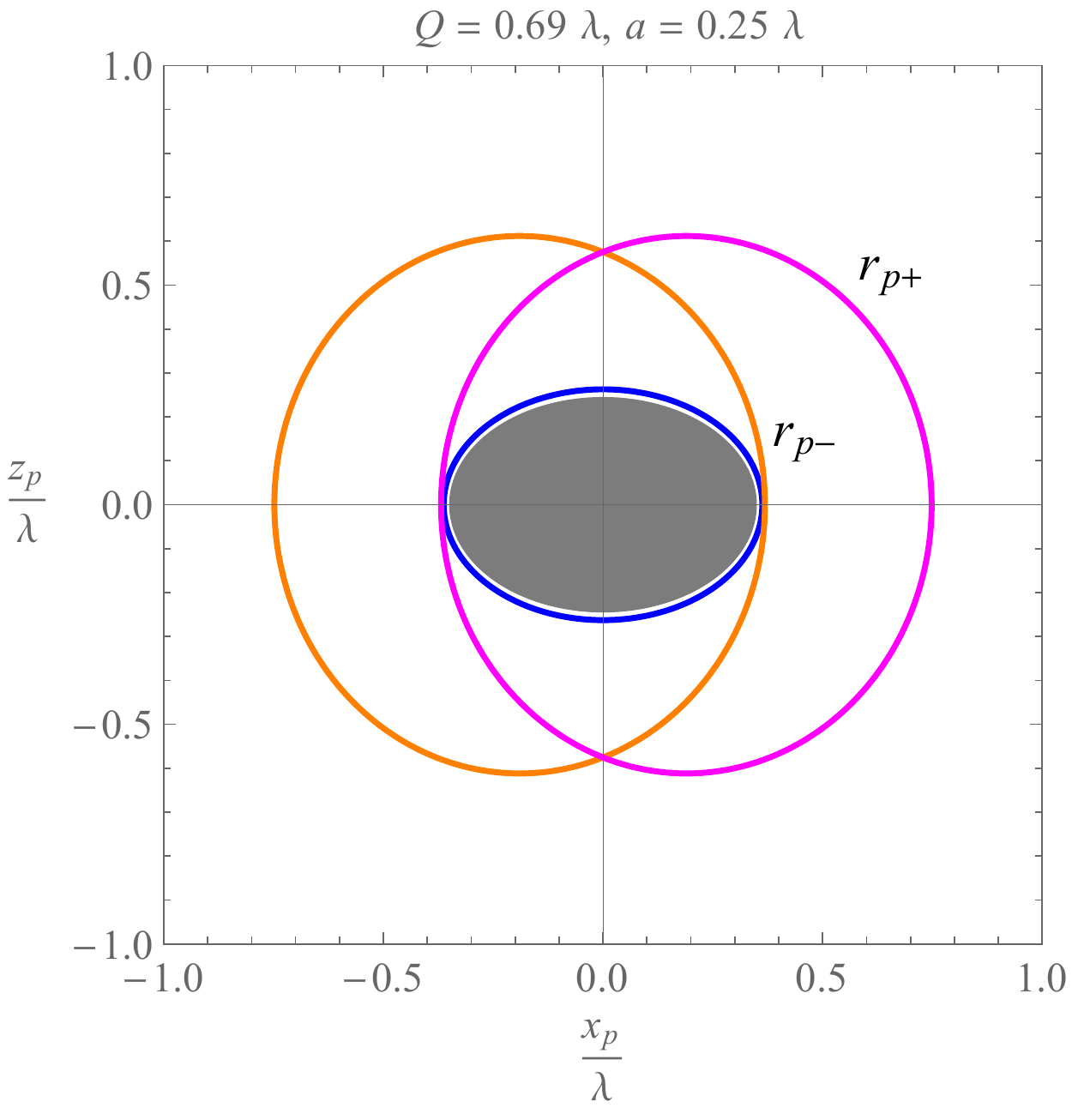}~(b)
    \includegraphics[width=5.4cm]{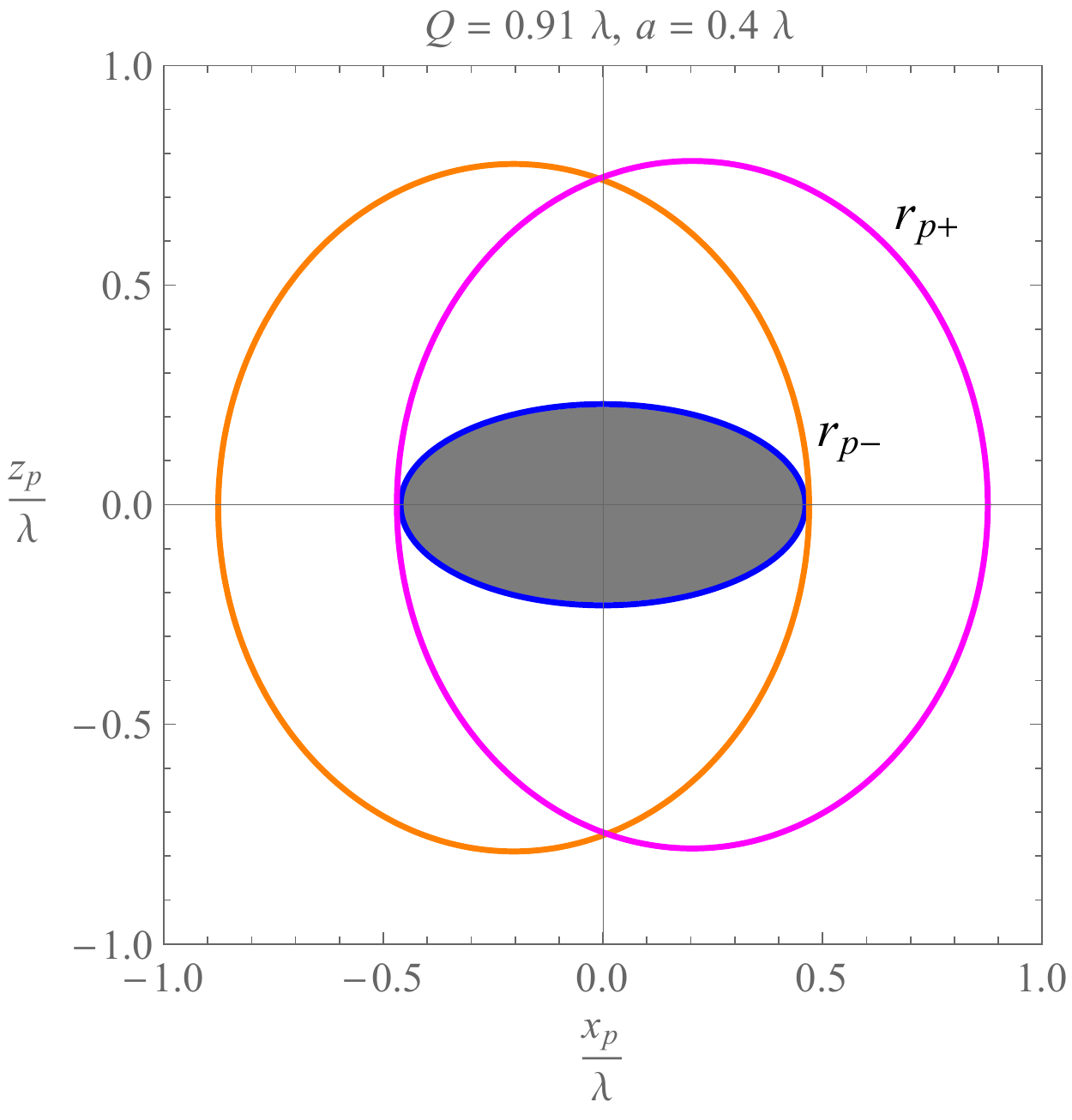}~(c)
    \caption{The shape of the photon regions ($r_{p-}<r<r_{p+}$), as viewed from the $y$ axis, plotted for $\lambda=10$, and different values of $Q$ and $a$. The event horizon $r_+$ occupies the central shaded region, and is encompassed by the rings $r_{p-}$ and $r_{p+}$. {The blue contour corresponds to the boundary of the inner ring $r_{p-}$, that coincides with the event horizon, only for the extremal black hole with $Q=Q_{\rm{ex}}$ and $r_{p-} = r_+ =  r_{\rm{ex}}$.}}
    \label{fig:PhotonSphere}
\end{figure}

Here, we apply the method of 
separation of variables in the Hamilton–Jacobi equation \cite{Carter:1968,Chandrasekhar:579245}. Accordingly, we write the Hamilton-Jacobi equation as
\begin{equation}\label{eq:H-J_0}
    \mathcal{H} = -\frac{\partial\mathcal{S}}{\partial\tau} = \frac{1}{2}g^{\alpha\beta} \frac{\partial\mathcal{S}}{\partial x^\alpha}\frac{\partial\mathcal{S}}{\partial x^\beta}= -\frac{1}{2}\mu^2,
\end{equation}
where $\mathcal{H}$, $\mathcal{S}$ and $\mu$, are respectively the canonical Hamiltonian, the Jacobi action and the rest mass of particles. Defining the four-momentum $p_\alpha= \partial\mathcal{S}/\partial x^\alpha$ ($p_\alpha p^\alpha = \mu^2$), the action can be separated by the Carter's method, as
\begin{equation}\label{eq:Jaction}
    \mathcal{S} = \frac{1}{2}\mu^2\tau-\tE t + \tL \phi + \mathcal{S}_r(r) + \mathcal{S}_\theta(\theta),
\end{equation}
in which $\mathcal{S}_r(r)\equiv p_r$, $\mathcal{S}_\theta(\theta)\equiv p_\theta$, and 
\begin{eqnarray}\label{eq:tE,tL}
    && \tE = - p_t = -\left(g_{tt}\frac{\ed t}{\ed\tau}+g_{t\phi}\frac{\ed\phi}{\ed\tau}\right),\label{eq:tE}\\
    && \tL = p_\phi = g_{\phi t}\frac{\ed t}{\ed\tau}+g_{\phi\phi}\frac{\ed\phi}{\ed\tau},\label{eq:tL}
\end{eqnarray}
are the constants of motion. Physically, $\tL$ is associated with the particles' angular momentum around the axis of symmetry. On the other hand, $\tE$ cannot be regarded as the energy of particles, because the spacetime under consideration is not asymptotically flat. Using Eqs.~\eqref{eq:Jaction}-\eqref{eq:tL} in the Hamilton-Jacobi equation \eqref{eq:H-J_0}, and applying the method of separation of variables for the case of mass-less particles (photons with $\mu=0$), the equations of motion are then given in terms of the four differential equations \cite{Chandrasekhar:579245}
\begin{eqnarray}\label{eq:geodesic_eqn}
    && \Sigma \frac{\ed t}{\ed\tau} = \frac{r^2+a^2}{\Delta}\left(
    \tE\left(r^2+a^2\right)-a\tL
    \right)-a\left(a\tE\sin^2\theta-\tL\right),\label{tdot}\\
   &&  \Sigma \frac{\ed r}{\ed \tau} = \pm\sqrt{\mathcal{R}(r)},\label{eq:rdot}\\
   &&  \Sigma \frac{\ed \theta}{\ed \tau} = \pm\sqrt{{\Theta}(\theta)},\label{eq:thetadot}\\
   && \Sigma\frac{\ed\phi}{\ed\tau} = \frac{a}{\Delta}\left(
    \tE\left(r^2+a^2\right)-a\tL
    \right)-\left(
    a\tE-\frac{\tL}{\sin^2\theta}
    \right),\label{eq:phidot}
\end{eqnarray}
defining
\begin{subequations}\label{eq:RTheta}
\begin{align}
    & \mathcal{R}(r)=\left(
    \left(r^2+a^2
    \right)\tE-a\tL
    \right)^2-\Delta\left(\mathcal{D}+
   \left(a\tE-\tL\right)^2
    \right),\label{eq:R(r)}\\
    & \Theta(\theta) = \mathcal{D} - \left(
    \frac{\tL^2}{\sin^2\theta} - a^2\tE^2
    \right)\cos^2\theta,\label{eq:Theta(theta)}
\end{align}
\end{subequations}
in which, $\mathcal{D}$ is the Carter's separation constant. The photon trajectories are therefore characterized by two dimension-less impact parameters
\begin{eqnarray}\label{eq:b,eta}
   && \tb = \frac{\tL}{\tE},\label{eq:b}\\
   && \teta = \frac{\mathcal{D}}{\tE^2}.
\end{eqnarray}
It is also common to use the generalized Carter's constant of motion $\K = \mathcal{D}+(a\tE-\tL)^2$ \cite{Chandrasekhar:579245}. As stated before, photon surfaces are those regions around the black hole, in which, the photons travel on unstable (critical) orbits. In this situation, the radial effective potential associated with the photon trajectories reaches its extremum at the corresponding critical distance $r_p$. Accordingly, Eq.~\eqref{eq:rdot} provides the conditions
\begin{subequations}\label{eq:CriticalR}
\begin{align}
    & \mathcal{R}(r_p) = 0,\label{eq:R(r)0}\\
    & \left.\frac{\partial\mathcal{R}(r)}{\partial r}\right|_{r=r_p} = 0,\label{eq:R(r)1}\\
    & \left.\frac{\partial^2\mathcal{R}(r)}{\partial r^2}\right|_{r=r_p} > 0\label{eq:R(r)2}.
\end{align}
\end{subequations}
The importance of the impact parameters $\tb$ and $\teta$ is their relevance to the fate of approaching photons to the black hole. In this regard, photons can either fall on unstable orbits, escape from, or captured by the black hole, respectively, if their associated impact parameters are equal, smaller, or larger than a critical value\footnote{See the discussion in Ref.~\cite{Fathi:2020} on the photon trajectories around the static CWBH.}. In fact, Eqs.~\eqref{eq:R(r)0} and \eqref{eq:R(r)1} result in the two equations
\begin{subequations}\label{eq:critical_eqns}
\begin{align}
 &  \left(a^2+r_p^2-a\, \tb_c\right)^2-\left(\teta_c+\left(a-\tb_c\right)^2\right)\Delta(r_p) =0,\\
 & 4 r_p^3+r_p\left(4 a^2-4 a\,\tb_c\right)-\left(\teta_c+\left(a-\tb_c\right)^2\right)\Delta'(r_p) =0,
\end{align}
\end{subequations}
in which, $\tb_c$ and $\teta_c$ are the critical values of these impact parameters, corresponding to the photons on unstable orbits at $r_p$. The above equations provide two pairs of $(\tb_c,\teta_c)$, that only the pair  
\begin{eqnarray}\label{eq:criticalb,eta}
    && \tb_c(r_p) = \frac{\left(a^2+r_p^2\right)\Delta'(r_p)-4 r_p \Delta(r_p)}{a \Delta'(r_p)},\label{eq:bc}\\
    && \teta_c(r_p) = \frac{r_p^2\left[
   8\Delta(r_p)\left(2 a^2-2\Delta(r_p)+r_p\Delta'(r_p)\right)-r_p^2\Delta'(r_p)^2
    \right]}{a^2 \Delta'(r_p)^2},\label{eq:etac}
\end{eqnarray}
satisfies the condition \eqref{eq:R(r)2}. Accordingly, the photons on unstable orbits are identified by $\teta_c = 0$, resulting in the two real positive values 
    \begin{eqnarray}
    && r_{p-} = \frac{Q}{\sqrt{1+\frac{4a^2}{\lambda^2}}} \sin\left(\frac{1}{2}
    \arcsin\left(
    \sqrt{\left(1-\frac{4a^2}{Q^2}\right)\left(1+\frac{4a^2}{\lambda^2}\right)}
    \right)
    \right),\label{eq:rp-}\\
    && r_{p+} = \frac{Q}{\sqrt{1+\frac{4a^2}{\lambda^2}}} \cos\left(\frac{1}{2}
    \arcsin\left(
    \sqrt{\left(1-\frac{4a^2}{Q^2}\right)\left(1+\frac{4a^2}{\lambda^2}\right)}
    \right)
    \right),\label{eq:rp+}
\end{eqnarray}
that implies $Q > 2a$ for $r_{p-}$ to exist. {In general, these solutions satisfy $r_+<r_{p_-}<r_{p+}<r_{++}$. However, to ensure the important condition $r_{p-}>\tilde{r}$, the cosmological component of the spacetime metric should satisfy $\tilde{\varepsilon}<\tilde{\varepsilon}_0$, where
\begin{equation}\label{eq:epsilon0}
   \tilde{\varepsilon}_0 = \frac{3 \left(4 a^2+4 \tilde{r} (\tilde{r}-3 \tilde{m})-Q^2\right)}{8 \tilde{r}^4},
\end{equation}
to obtain which, we have used Eq.~\eqref{eq:rp-} and the expression in Eq.~\eqref{par1}. Note that, Eqs.~\eqref{eq:horizons_new_rp} and \eqref{eq:rp-} imply that the inner photon ring is identified with the event horizon under the critical condition $Q=Q_{\rm{ex}}$, that corresponds to the extremal black hole, for which $r_{p-}=r_{+}=r_{\rm{ex}}$.}
Furthermore, the photon ring radii should also respect the condition $\Theta(\theta)\geq 0$ \cite{Chandrasekhar:579245}, for which, Eq.~\eqref{eq:Theta(theta)} yields $\teta_c\geq\tb_c^2\cot^2\theta-a^2\cos^2\theta$ on the photon surface, or by using Eqs.~\eqref{eq:bc} and \eqref{eq:etac},
\begin{equation}\label{eq:Theta(theta)>0}
    4r_p^2 \Delta(r_p)\left(\frac{\Delta(r_p)}{\sin^2\theta}-9\right)
    \leq
    \left(9+2r_p^2+9\cos(2\theta)\right)\frac{\Delta'(r_p)}{\sin^2\theta}\left(r_p\Delta(r_p)-\frac{\Delta'(r_p)}{16}\right).
\end{equation}
Applying the Cartesian coordinates in Eq.~\eqref{eq:Cartesian}, $r_{p-}$ and $r_{p+}$ and the corresponding photon regions ($r_{p-}<r<r_{p+}$), have been plotted in Fig.~\ref{fig:PhotonSphere}, together with the event horizon, for some definite values of the black hole parameters. As it is observed from the figures, faster spinning RCWBH can produce larger photon regions.
In fact, the shape of the photon regions informs about that of the black hole shadow, since the shadow is confined to the photon surfaces. Those photons on unstable orbits that can reach the distant observers, can create an image of the outer regions of the event horizon. Such photons are therefore responsible, for example, for the image obtained from M87* \cite{Akiyama:2019}. 

To proceed with the determination of the shadow of the RCWBH, we should bear in mind that this spacetime is not asymptotically flat. We therefore cannot consider an observer at infinity, as is traditionally done in Refs.~\cite{Bardeen:1973a,Bardeen:1973b} and in Ref.~\cite{Vazquez:2004}. Therefore, instead of using the celestial coordinates in the sky of an observer at infinity, we locate an observer at the coordinate position $(r_o,\theta_o)$, which is characterized by the orthonormal tetrad $\bm{e}_{\{A\}}$, selected as
\begin{eqnarray}\label{eq:tetrad}
&& \bm{e}_0 = \left.\frac{\left(\Sigma+a^2\sin^2\theta\right)\partial_t+a\,\partial_\phi}{\sqrt{\Sigma\, \Delta}}\right|_{(r_o,\theta_o)},\label{eq:e0}\\
&& \bm{e}_1 = \left.\sqrt{\frac{1}{\Sigma}}\, \partial\theta\right|_{(r_o,\theta_o)},\label{eq:e1}\\
&& \bm{e}_2 = \left.-\frac{\left(\partial_\phi+a \sin^2\theta\,\partial_t\right)}{\sqrt{\Sigma } \sin\theta}\right|_{(r_o,\theta_o)},\label{eq:e2}\\
&& \bm{e}_3 = \left.-\sqrt{\frac{\Delta}{\Sigma}}\, \partial r\right|_{(r_o,\theta_o)},\label{eq:e3}
\end{eqnarray}
that satisfy ${{e}_A}^\alpha {{e}^B}_\alpha = \delta^B_A$. This method, introduced in Ref.~\cite{Grenzebach:2014}, makes it possible to calculate the celestial coordinates in general spacetimes with cosmological constituents (also see Ref.~\cite{Grenzebach:2016} for a review). In the above set of tetrads, the time-like vector $\bm{e}_0$ is supposed to be the velocity four-vector of the selected observer. Furthermore, $\bm{e}_3$ is set to point towards the black hole and $\bm{e}_0\pm\bm{e}_3$ is the generator of the principal null congruence. This way, a linear combination of $\bm{e}_{\{A\}}$ is tangent to the light ray $\bm\ell(\tau) = \left(t(\tau),r(\tau),\theta(\tau),\phi(\tau)\right)$, which is sent from the black hole to the past. The tangent to this light ray can be parameterized in two ways as
\begin{eqnarray}\label{eq:elldot}
 &&   \frac{\ed\bm{\ell}}{\ed\tau} = \frac{\ed t}{\ed\tau} \, \partial_t + \frac{\ed r}{\ed \tau}\,\partial_r+\frac{\ed\theta}{\ed\tau}\, \partial_\theta + \frac{\ed\phi}{\ed\tau}\, \partial_\phi,\\
 &&   \frac{\ed\bm{\ell}}{\ed\tau} = \mathfrak{c}\left(-\bm{e}_0+\sin\vartheta\cos\psi\,\bm{e}_1 + \sin\vartheta\sin\psi\,\bm{e}_2 + \cos\vartheta\,\bm{e}_3\right),
\end{eqnarray}
in which, $\vartheta$ and $\psi$ are newly defined celestial coordinates in the observer's sky and 
\begin{equation}\label{eq:c}
    \mathfrak{c} = \bm{g}\left(\bm{\ell},\bm{e}_0\right) = \frac{a \tL-\left(\Sigma+a^2\sin^2\theta\right)\tE}{\sqrt{\Sigma\,\Delta}}.
\end{equation}
It can be easily noticed that $\vartheta = 0$ points, directly, to the black hole. Since the boundary curve of the shadow is generated by those light rays that come onto the critical (unstable) null geodesics at the radial distance $r_p$, this region therefore corresponds to the critical impact parameters $\tb_c$ and $\teta_c$, given in Eqs.~\eqref{eq:bc} and \eqref{eq:etac}. The corresponding celestial coordinates $(\psi_p,\vartheta_p)$ at this distance, for the observer located at $(r_o,\theta_o)$, have been derived as \cite{Grenzebach:2014}
\begin{eqnarray}\label{eq:vartheta,psi}
    && \mathcal{P}(r_p,\theta_o) \vcentcolon= \sin\psi_p = \frac{\tb_c(r_p)+a\cos^2\theta_o-a}{\sqrt{\teta_c(r_p)}\,\sin\theta_o},\label{eq:psi_p}\\
     &&  \mathcal{T}(r_p,r_o) \vcentcolon= \sin\vartheta_p = \frac{\sqrt{\Delta(r_o)\,\teta_c(r_p)}}{r_o^2-a (\tb_c(r_p)-a)}.\label{eq:vartheta_p}
\end{eqnarray}
Accordingly, the $\vartheta$ coordinate has its maximum and minimum values, respectively, for $\psi_p=-\pi/2$ and $\psi_p=\pi/2$. This helps us obtaining the corresponding values of $r_p$ for each of the cases. Applying Eq.~\eqref{eq:psi_p}, these conditions result in the equation
\begin{equation}\label{eq:rpmax,minEQ}
    \Sigma(r_p,\theta_o)\Delta'(r_p)-4r_p\Delta(r_p) = \mp r_p\sin\theta_o  \sqrt{16\left(a^2-\Delta(r_p)\right)\Delta(r_p)+8r_p\Delta(r_p)\Delta'(r_p)-r_p^2\Delta'(r_p)^2},
\end{equation}
where $\Sigma(r_p,\theta_o) = r_p^2+a^2\cos^2\theta_o$. This equation is of fourth order in $r_p$, and has the positive solutions
\begin{eqnarray}\label{eq:rpmax,minSOL}
 && r_{p{\min}} = \sqrt{\frac{\bar{c}_2}{\bar{c}_1}}\,\sin\left(\frac{1}{2}\arcsin\left(
 \frac{2\sqrt{\bar{c}_3/\bar{c}_1}}{\bar{c}_2/\bar{c}_1}
 \right)\right),\label{rpmax}\\
 && r_{p{\max}} = \sqrt{\frac{\bar{c}_2}{\bar{c}_1}}\,\cos\left(\frac{1}{2}\arcsin\left(
 \frac{2\sqrt{\bar{c}_3/\bar{c}_1}}{\bar{c}_2/\bar{c}_1}
 \right)\right),\label{rpmin}
\end{eqnarray}
in which
\begin{subequations}\label{cb1,cb2,cb3}
\begin{align}
    & \bar{c}_1 = 4 \left(4 a^2 \cot ^2{\theta_o} \left(a^2 \cos ^2{\theta_o}+\lambda ^2\right)-\lambda ^2 \left(4 a^2+\lambda ^2\right)+\lambda ^4 \csc ^2{\theta _o}\right),\\
    &  \bar{c}_2 = 2 \lambda ^4 \csc ^2{\theta_o} \left(-8 a^2+Q^2 \cos (2 {\theta _o})+Q^2\right)+8 a^2 \lambda ^2 \cot ^2{\theta _o} \left(a^2 \cos (2 {\theta_o})-3 a^2+\lambda ^2+Q^2\right),\\
    & \bar{c}_3 = \frac{1}{2} \lambda ^4 \csc ^2{\theta _o} \left(a^4 \cos (4 {\theta _o})+\left(Q^4-12 a^4\right) \cos (2 {\theta _o})+19 a^4-8 a^2 Q^2+Q^4\right).
\end{align}
\end{subequations}
The above values are indeed those radii where the boundary of the photon region intersects with the cone $\theta=\theta_o$. When $a=0$, we have $r_{p{\max}} = r_{p{\min}} = Q/\sqrt{2}$, which is the radius of the unstable photon orbits around a static CWBH, and it does not depend on $\lambda$ \cite{Fathi:2020}. This unique value corresponds to the constant value of $\vartheta_p = \pi/2$ for $a=0$. The shadow of the static black hole is therefore circular.

\begin{figure}
    \centering
    \includegraphics[width=7cm]{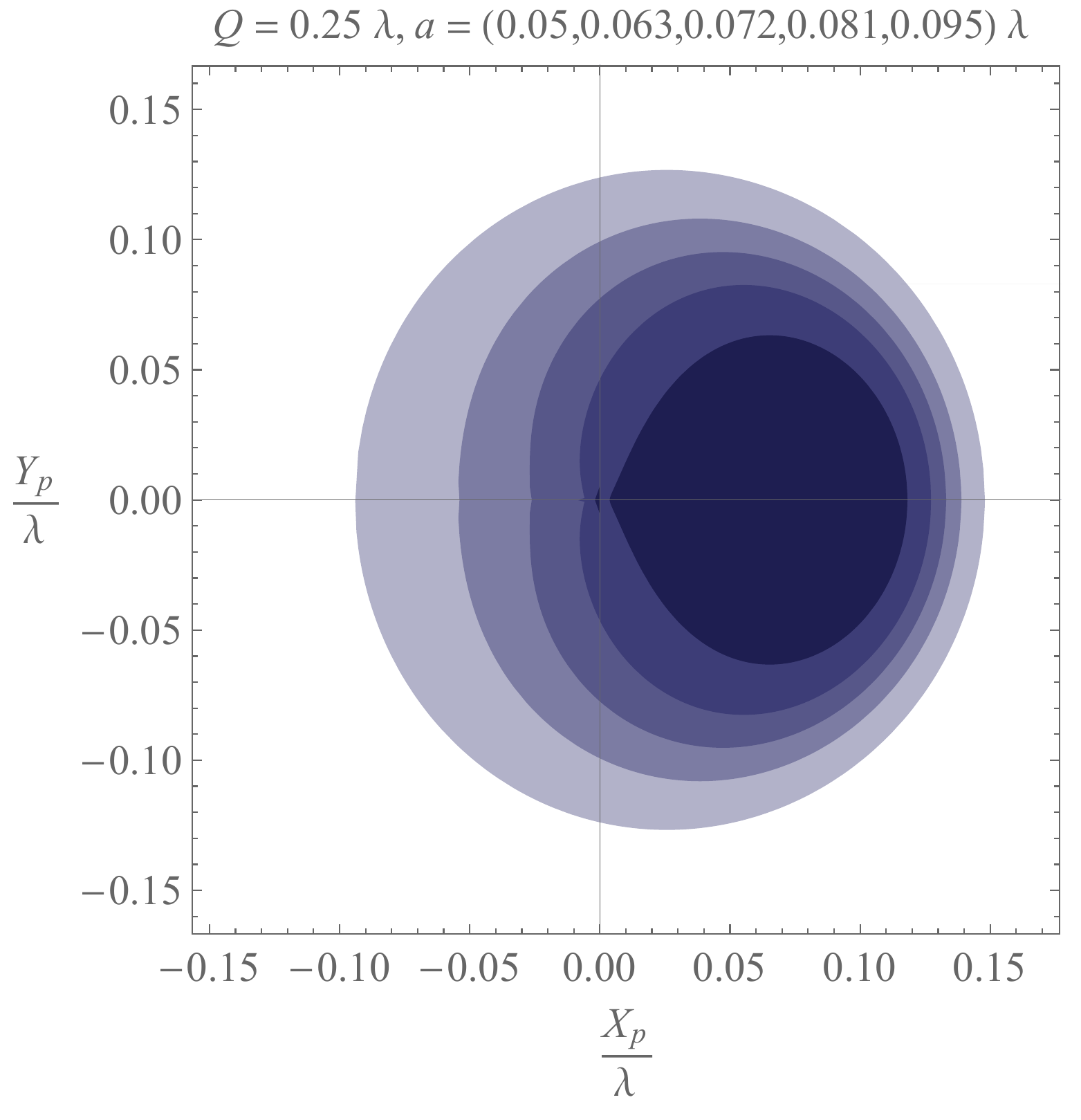}~(a)
    \includegraphics[width=7cm]{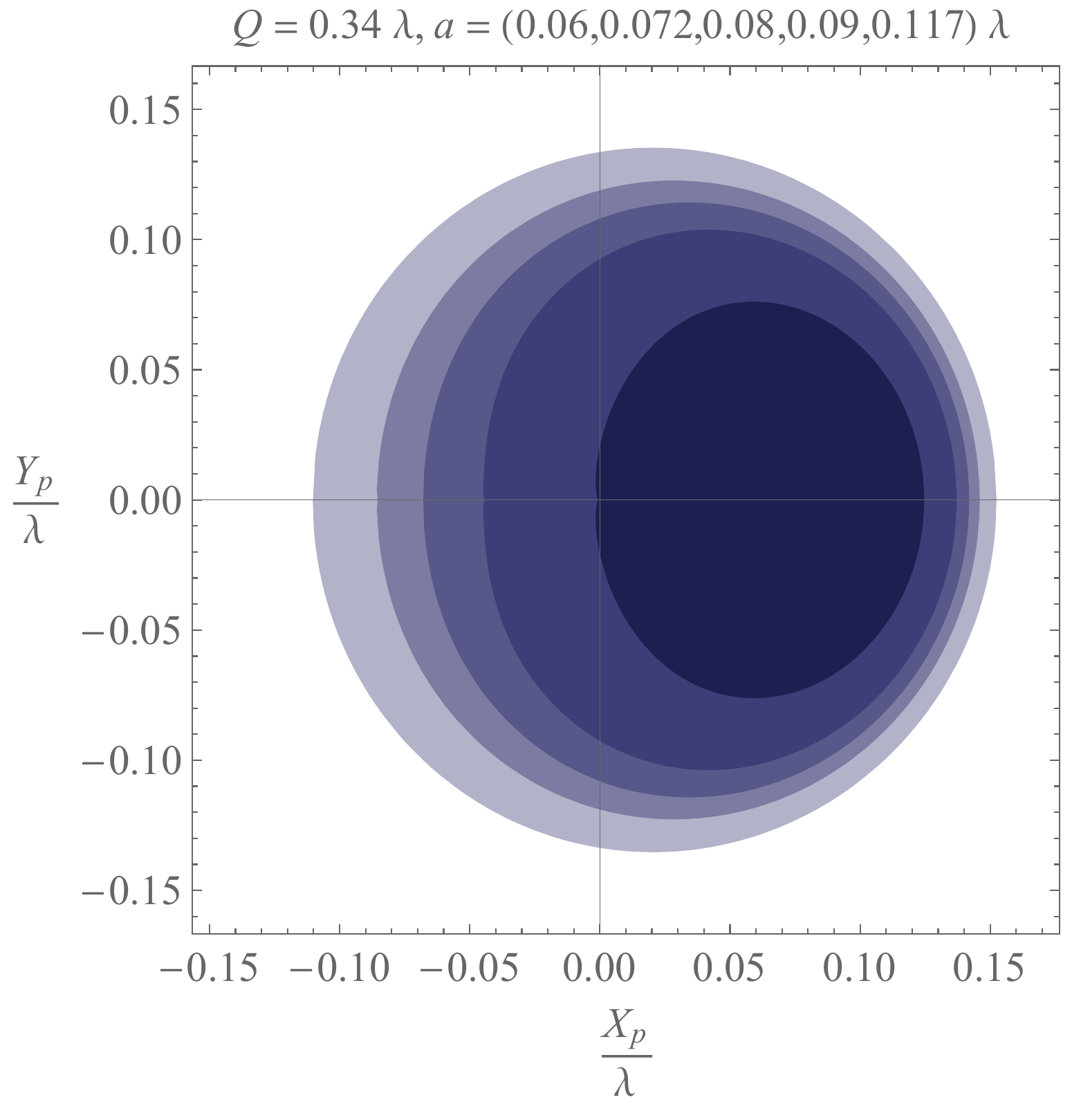}~(b)
    \includegraphics[width=7cm]{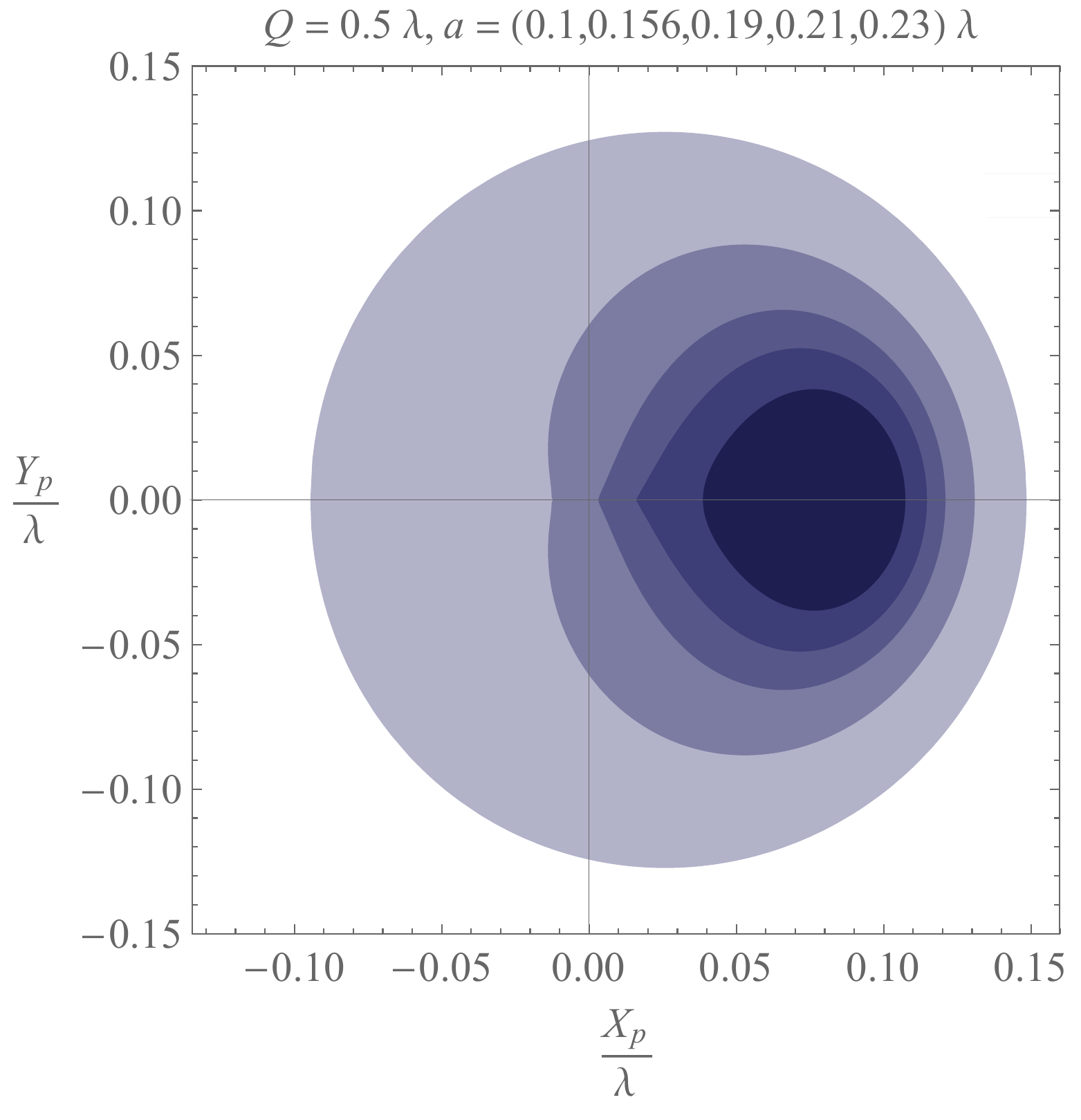}~(c)
    \includegraphics[width=7cm]{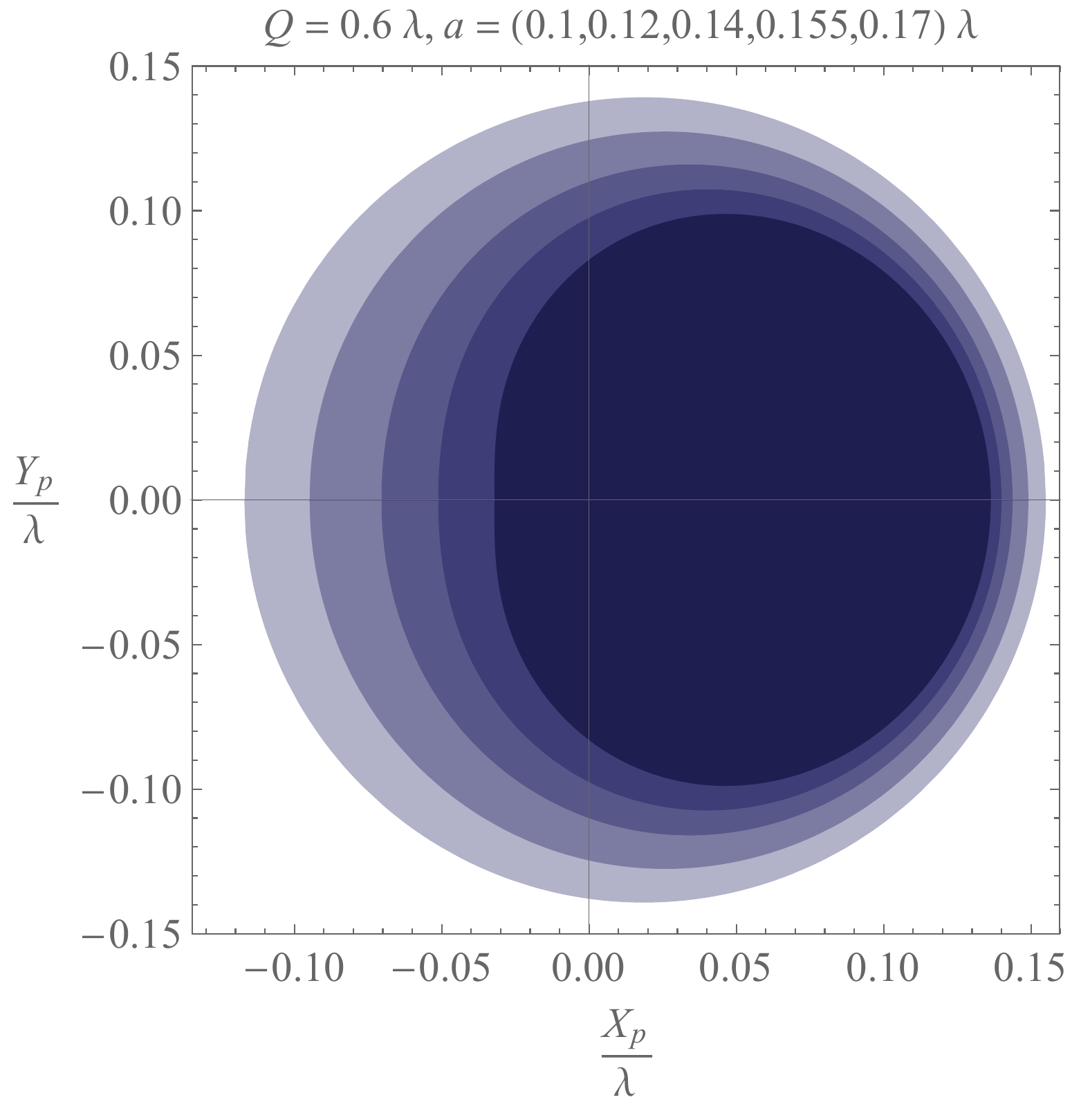}~(d)
    \includegraphics[width=7cm]{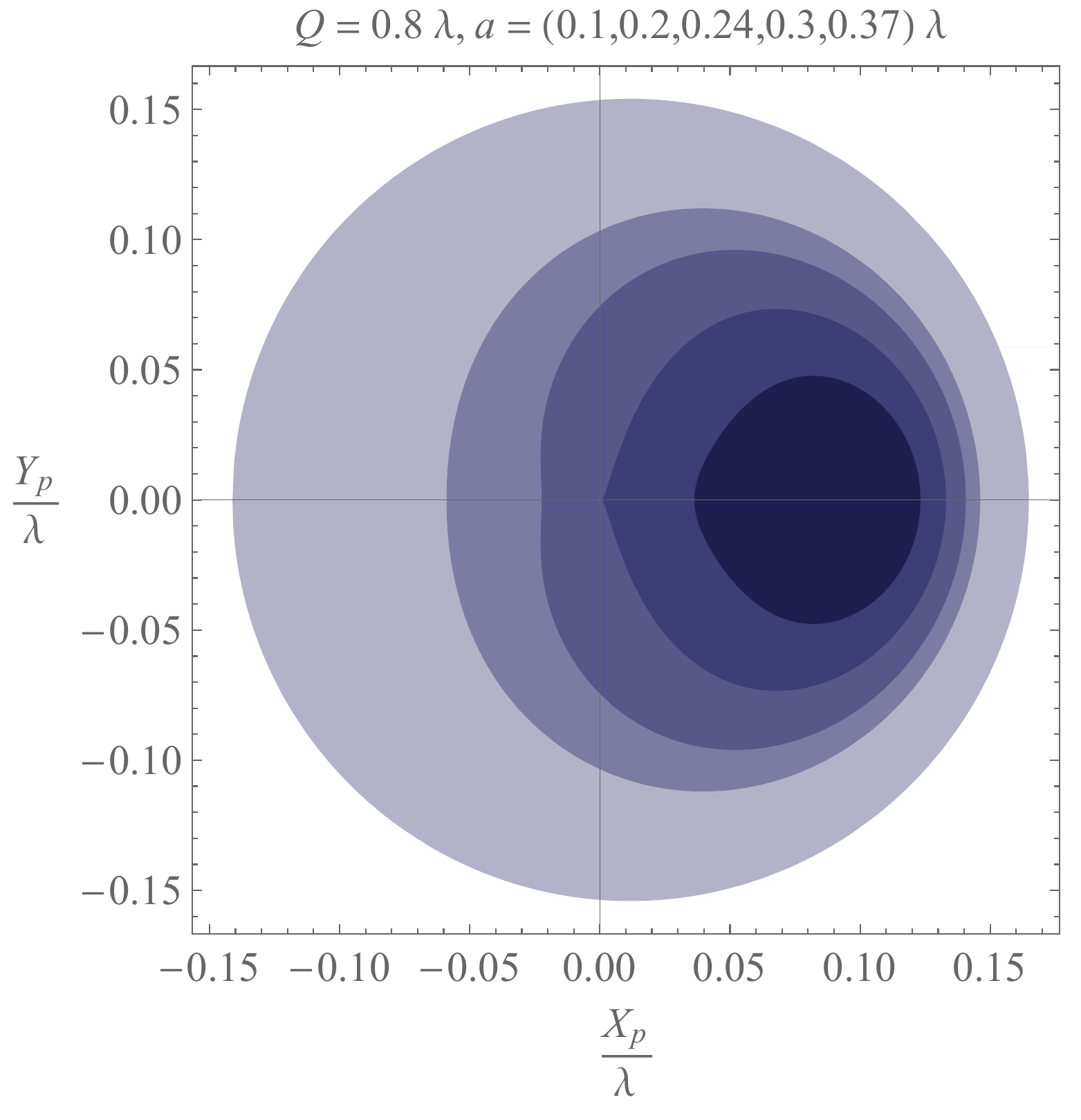}~(e)
    \includegraphics[width=7cm]{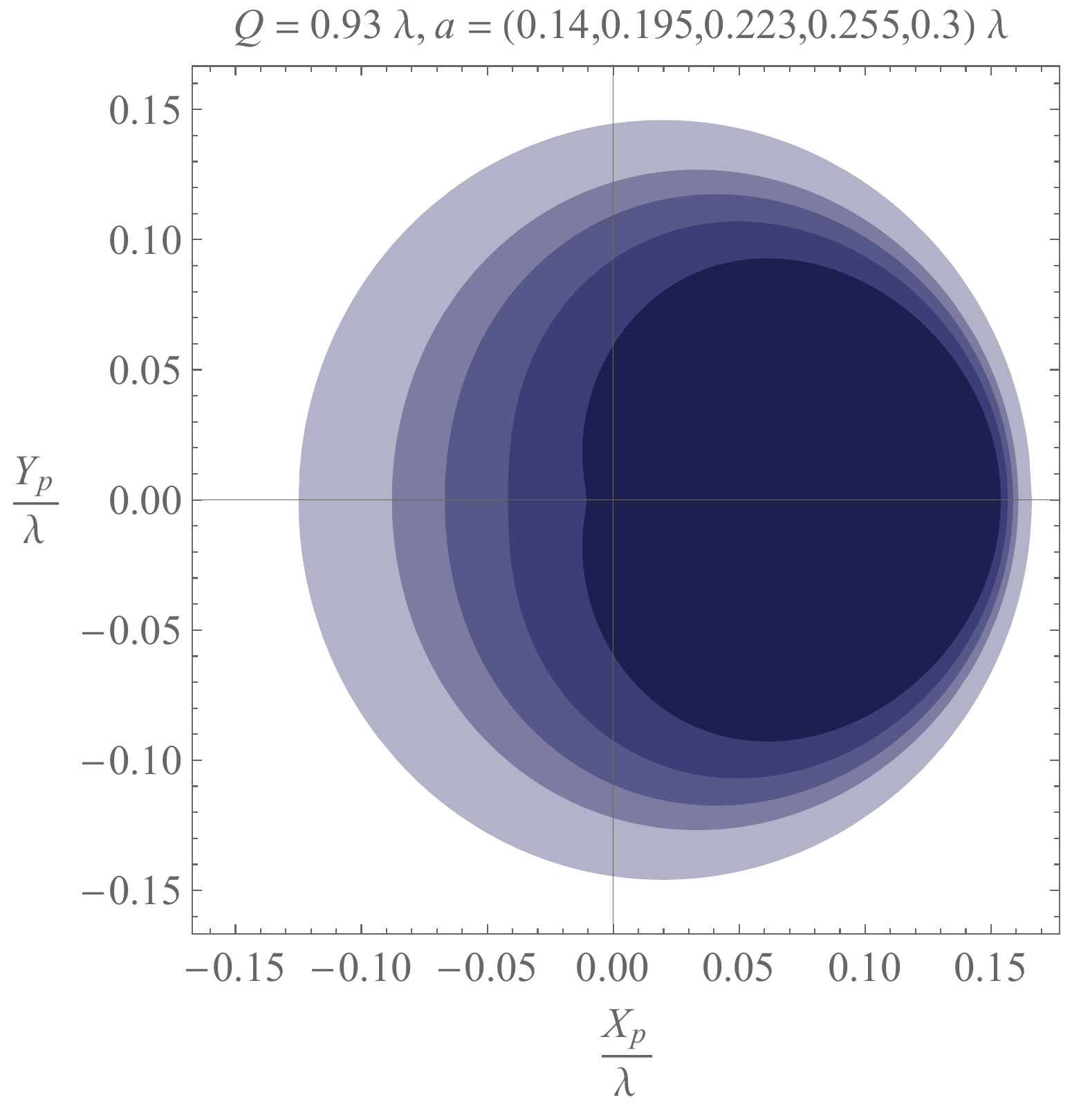}~(f)
    \caption{The shadow of the RCWBH for $a\neq0$, plotted for $\lambda=10$ and $r_o = 5 \lambda$. Each diagram corresponds to a fixed value of $Q$ and five values of $a$, which are sorted from the largest to the smallest parametric curve. The figures have been scaled in order to have an approximate circular shape for the smallest $a$ in each of the diagrams.}
    \label{fig:shadow_rotating}
\end{figure}

Now, getting back to the case of $a\neq0$, the two-dimensional Cartesian coordinates for the chosen observer of the velocity four-vector $\bm{e}_0$, are now obtained by applying the stereographic projection of the celestial sphere $(\psi_p,\vartheta_p)$, onto a plane. This provides the coordinates \cite{Grenzebach:2014}
\begin{eqnarray}\label{eq:Xp,Yp}
 &&   X_p = -2 \tan\left(\frac{\vartheta_p}{2}\right)\sin\psi_p,\label{eq:Xp}\\
&&   Y_p = -2 \tan\left(\frac{\vartheta_p}{2}\right)\cos\psi_p.\label{eq:Yp}
\end{eqnarray}
The case of $\theta_o = \pi/2$ corresponds to the equatorial plane of view for the observer. Taking this into account, in Fig.~\ref{fig:shadow_rotating}, we have used the above coordinates to obtain the shadow of the RCWBH, for different values of electric charge and spin parameter. The curves use $r_p$ as their parameter. 
In general, and for a given spin parameter, the size of the shadow increases by increasing the electric charge, and as it is inferred from the figures, by raising the black hole's spin, the shadow shrinks and tends to the positive part of the coordinate plane. It can also be noted that the shadow becomes oblate toward the $X_p = 0$ axis from the negative sector of the coordinate plane, whereas it is sharp toward that, from the positive sector. However, the amount of such deformations cannot be inferred directly from the $Q/a$ fraction, and is indeed a consequence of the spacetime's response to the changes in the electric charge. In fact, the size and the deformation of the shadow casts of black holes have been used to estimate their dynamical properties. In the forthcoming subsection, using a specific geometric method, we relate the angular size of a deformed shadow to the physical characteristics of the black hole. {However, before proceeding with that, it is of worth to make a comparison between the formation and the evolution of the shadows of the Kerr-Newman-de Sitter black hole (KNdSBH) and the RCWBH. This way, we will be able to gain some visualizations for the differences between the general relativistic spacetimes and that of the RCWBH. The line element associated with the KNdSBH is given by \cite{Griffiths:2009}
\begin{multline}\label{eq:KNdS}
\ed {s}_{\rm{kn}}^2 =-\frac{\Delta_r-a^2 \Delta_\theta\sin^2\theta}{\Sigma}\,\ed t^2 +  \frac{\Sigma}{\Delta_r}\,\ed r^2 + \frac{\Sigma}{\Delta_\theta}\,\ed\theta^2+\frac{2}{\Sigma}\left[
\Delta_r a \sin^2\theta - a \Delta_\theta\sin^2\theta \left(\Sigma+a^2\sin^2\theta\right)
\right]\ed t\,\ed\phi\\
+\frac{1}{\Sigma}\left[
\left(\Sigma+a^2\sin^2\theta\right)^2\Delta_\theta\sin^2\theta-\Delta_r a^2\sin^4\theta
\right]\ed\phi^2,
\end{multline}
in which, the new functions
\begin{subequations}
\begin{align}
    & \Delta_r = r^2-2M_0r+a^2+Q_0^2+\frac{R_0 r^2}{12}\left(a^2+r^2\right),\\
    & \Delta_\theta = 1+ \frac{R_0 a^2\cos^2\theta}{12},
\end{align}
\end{subequations}
with $R_0 \equiv 4\Lambda$, associate with a massive object of mass $M_0$ and charge $Q_0$. In Ref.~\cite{Grenzebach:2014}, the shadow of the KNdSBH has been calculated, using the celestial coordinates 
\begin{eqnarray}\label{eq:thetapsi_KN}
 && \sin\psi_{p}^{\rm{kn}} =   \frac{1}{\sqrt{\Delta_\theta(\theta_o)}}\sin\psi,\\
 && \sin\vartheta_p^{\rm{kn}} = \sin\vartheta_p,
\end{eqnarray}
which are here given in terms of those in Eqs.~\eqref{eq:psi_p} and \eqref{eq:vartheta_p}, and the same impact parameters in Eqs.~\eqref{eq:bc} and \eqref{eq:etac}. Defining $\overset{\bullet}{\mathfrak{q}}$, as the dimension-less version of a black hole quantity $\mathfrak{q}$, re-scaled by the corresponding black hole mass parameters\footnote{{For example, $\overset{\bullet}{a}$ means $a/\tilde{m}$ for the RCWBH and  $a/M_0$ for the KNdSBH, and etc. In particular, $\overset{\bullet}{\tilde{\varepsilon}} = \tilde{\varepsilon}/\tilde{m}^2$ and $\overset{\bullet}{R}_0 = R_0/M_0^2$.}}, in Fig.~\ref{fig:KNdSBH}, we have applied the Cartesian coordinates \eqref{eq:Xp} and \eqref{eq:Yp}, to compare the shadows of the KNdSBH and the RCWBH, for definite values of the black hole parameters, and several values of the spin parameter. Having in mind the re-scaling of the coordinates, it is seen from the diagrams that the RCWBH has larger shadows than the KNdSBH, respecting their black hole masses. {Additionally, the increase in the spin parameter, makes the shadow of the KNdSBH to shrink outside the larger ones, whereas for the RCWBH, this happens inside of those.}
\begin{figure}[t]
    \centering
    \includegraphics[width=8cm]{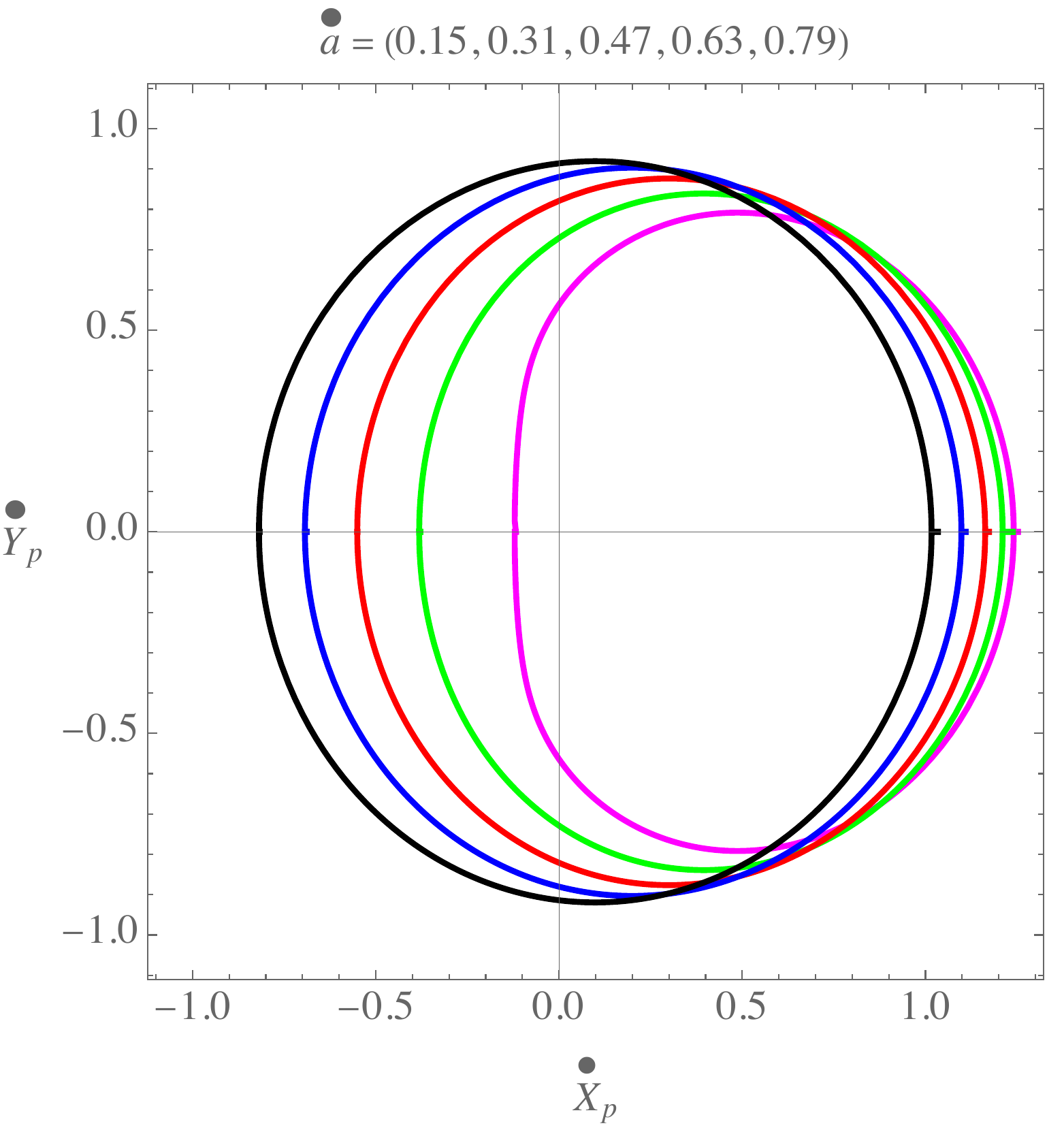} {(a)}
    \includegraphics[width=8cm]{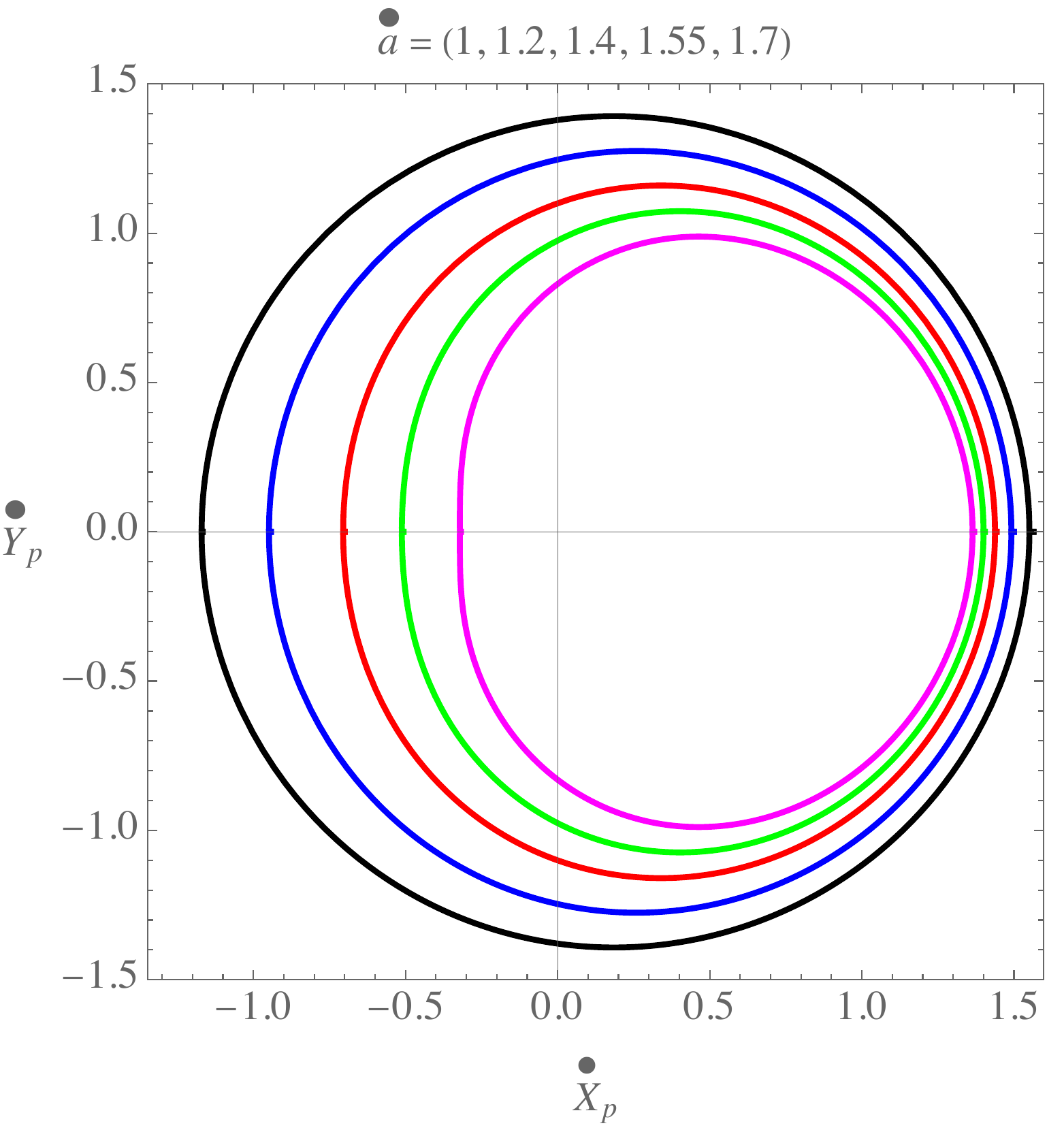} {(b)}
    \caption{{Shadows of {{(a)}} the KNdSBH 
    and {{(b)}} the RCWBH 
    plotted for $\overset{\bullet}{Q} = 6$, $\overset{\bullet}{r}_o = 5$ and $\theta_o = \pi/2$, for both of the black holes. The cosmological terms have been considered as $\overset{\bullet}{R}_0 = \overset{\bullet}{\tilde{\varepsilon}} = 10^{-3}$. For the case of RCWBH, this value corresponds to $\lambda = 10$, if $\overset{\bullet}{\tilde{r}}\approx 6.93$. The values of the re-scaled spin parameter $\overset{\bullet}{a}$, have been sorted from the less oblate to the most oblate parametric curves. As it is observed, respecting the corresponding black hole masses, the shadow of the RCWBH is greater in size.}}
    \label{fig:KNdSBH}
\end{figure}}

\subsection{The angular size of the shadow}

\begin{figure}[t]
    \centering
    \includegraphics[width=16cm]{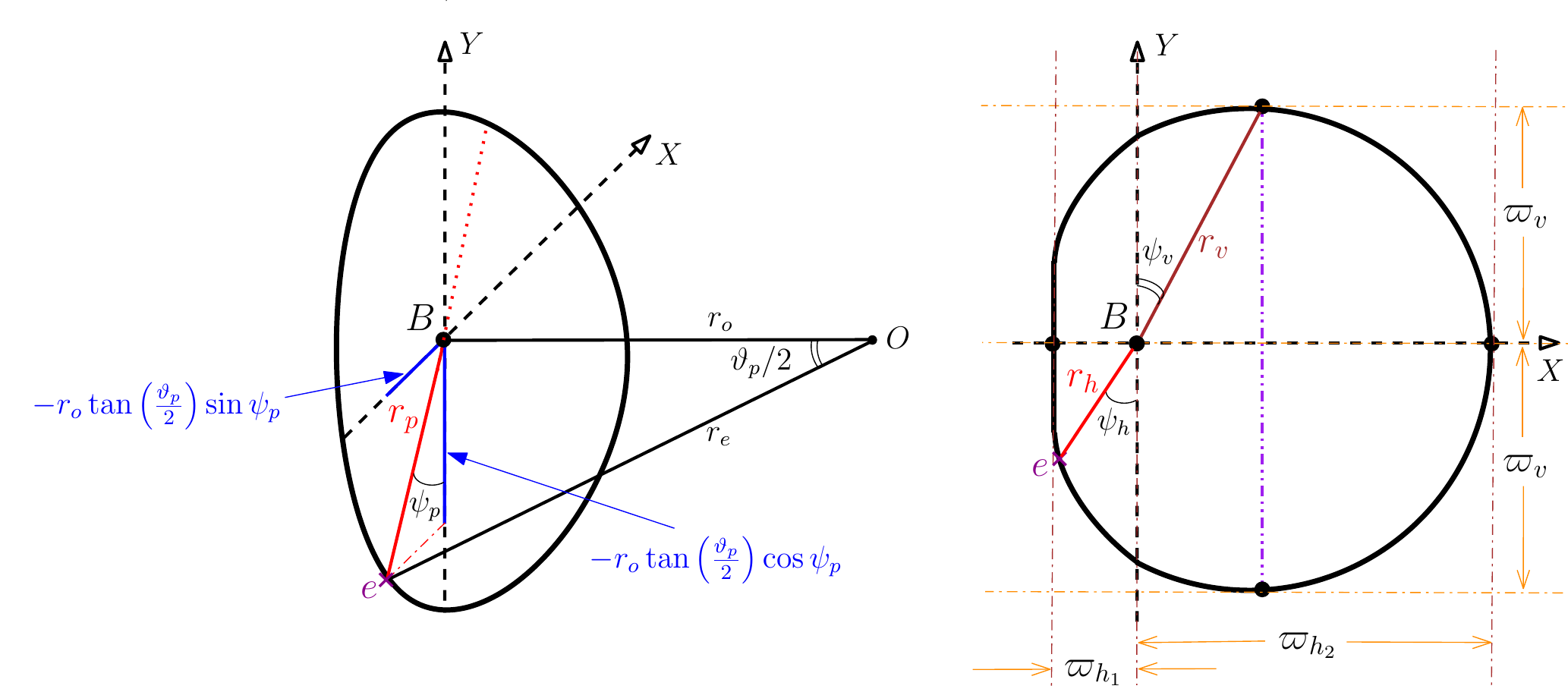}
    \caption{Characterization of an oblate shadow in terms of the angular diameters. The celestial coordinates $(\psi_p,\vartheta_p)$ are transformed to the Cartesian coordinates $(X,Y)$, as indicated in the left panel. The black hole at point $B$ is located at the center of the Cartesian coordinates. The observer at point $O$ and at the distance $r_o$ from the black hole, observes the event $e$ on the shadow. In the right panel, and following the method given in Ref.~\cite{Grenzebach:2015}, the angular diameters corresponding to the mentioned celestial coordinates, are now expressed in terms of three angular radii $\varpi_{h_1}$, $\varpi_{h_2}$ and $\varpi_v$, according to the symmetry with respect to the  $X$ coordinate.}
    \label{fig:M_shadow_cord}
\end{figure}

The celestial coordinates defined in Eqs.~\eqref{eq:psi_p} and \eqref{eq:vartheta_p} can be used in order to calculate the angular diameters of the shadow. As shown in the left panel of Fig.~\ref{fig:M_shadow_cord}, these angular diameters are expressed in terms of the celestial coordinates $(\psi_p,\vartheta_p)$. 
Following the methods introduced in Ref.~\cite{Grenzebach:2015}, we replace these angular diameters by
\begin{eqnarray}\label{eq:angula_radii_0}
  && \delta_h = \varpi_{h_1}+\varpi_{h_2},\label{eq:deltah}\\
  &&  \delta_v = 2\varpi_v,\label{eq:deltav}
\end{eqnarray}
which are defined in terms of the three angular radii $\varpi_{h_1}$, $\varpi_{h_2}$ and $\varpi_v$, that obey the properties (see the right panel of Fig.~\ref{fig:M_shadow_cord}) 
\begin{eqnarray}\label{eq:h,vproperties}
&& \sin{\varpi_{h_i}} =  \sin{\vartheta_{h_i}} \sin{\psi_{h_i}},~~~~(i=1,2),\\
&& \sin{\varpi_v} =  \sin{\vartheta_v} \cos{\psi_v},
\end{eqnarray}
or from Eqs.~\eqref{eq:psi_p} and \eqref{eq:vartheta_p},
\begin{eqnarray}\label{eq:h,vproperties_1}
&& \sin{\varpi_{h_i}} =  \mathcal{T}(r_{h_i},r_o) \mathcal{P}(r_{h_i},\theta_o),\label{eq:sinh}\\
&& \sin{\varpi_v} =  \mathcal{T}(r_v,r_o)\sqrt{1-\mathcal{P}^2(r_v,\theta_o)}.\label{eq:sinv}
\end{eqnarray}
Once again we restrict ourselves to the equatorial plane by letting $\theta_o = \pi/2$, so that the horizontal angular diameter corresponds to $\psi_{h} = \pm \pi/2$, and we need to solve $\mathcal{P}^2(r_h,\pi/2)=1$ for this case. Using Eq.~\eqref{eq:psi_p} together with Eqs.~\eqref{eq:bc} and \eqref{eq:etac}, this condition provides an equation of fourth order in $r_h$, which has the positive solutions
\begin{eqnarray}\label{eq:rh1-rh2}
    && r_{h_1} = \sqrt{\bar{C}_1} \sin\left(
    \frac{1}{2}\arcsin\left(\frac{2\sqrt{\bar{C}_2}}{\bar{C}_1}\right)
    \right),\label{eq:rh1}\\
    && r_{h_2} = \sqrt{\bar{C}_1} \cos\left(
    \frac{1}{2}\arcsin\left(\frac{2\sqrt{\bar{C}_2}}{\bar{C}_1}\right)
    \right),\label{eq:rh1}
\end{eqnarray}
where
\begin{subequations}\label{eq:C1-C2}
\begin{align}
& \bar{C}_1 = \frac{Q^2-2a^2}{2\left(\frac{a^2}{\lambda^2}\right)+1},\\
& \bar{C}_2 = \frac{2a^4-\frac{3}{2}a^2\left(a^2+\frac{1}{4}\right)}{\left(\frac{a^2}{\lambda^2}\right)+1}.
\end{align}
\end{subequations}
The horizontal angular radii are then calculated by evaluating the product $\mathcal{T}(r_{h_{1,2}},r_o) \mathcal{P}(r_{h_{1,2}},\pi/2)$, and from there, $\delta_h$ can be evaluated using Eq.~\eqref{eq:deltah}. To calculate the vertical angular radius, we take into account the fact that this radius corresponds to those points on the shadow's boundary, where the tangent to the curve $\mathcal{C}(r_v,r_o,\theta_o)\equiv\sin^2\varpi_v$ vanishes. Accordingly, we encounter the equation $\partial_{r_v}\mathcal{C}(r_v,r_o,\pi/2) = 0$. Considering this condition and applying Eq.~\eqref{eq:sinv} together with Eqs.~\eqref{eq:psi_p} and \eqref{eq:vartheta_p}, we obtain the unique positive value
\begin{equation}\label{eq:rv}
    r_v = r_o\sqrt{\frac{Q^2-2a^2}{2\left(a^2+r_o^2\right)}},
\end{equation}
which evaluates the vertical angular radii as
\begin{equation}\label{eq:C(rv)}
    \sin^2\varpi_v = \frac{\left(Q^2-2a^2\right) \left(\lambda ^2 \left(4 a^2-Q^2+4 r_o^2\right)-4 r_o^4\right)}{8 a^4 \lambda ^2+a^2 \left(8 r_o^2 \left(\lambda ^2+r_o^2\right)-2 \lambda ^2 Q^2\right)+4 r_o^4 \left(\lambda ^2-Q^2\right)}.
\end{equation}
The vertical angular diameter $\delta_v$, can be therefore calculated using Eq.~\eqref{eq:deltav}. In Fig.~\ref{fig:deltah,deltav}, the behaviors of $\delta_h$ and $\delta_v$ have been plotted in terms of changes in $Q$ and $a$.
\begin{figure}
    \centering
    \includegraphics[width=7cm]{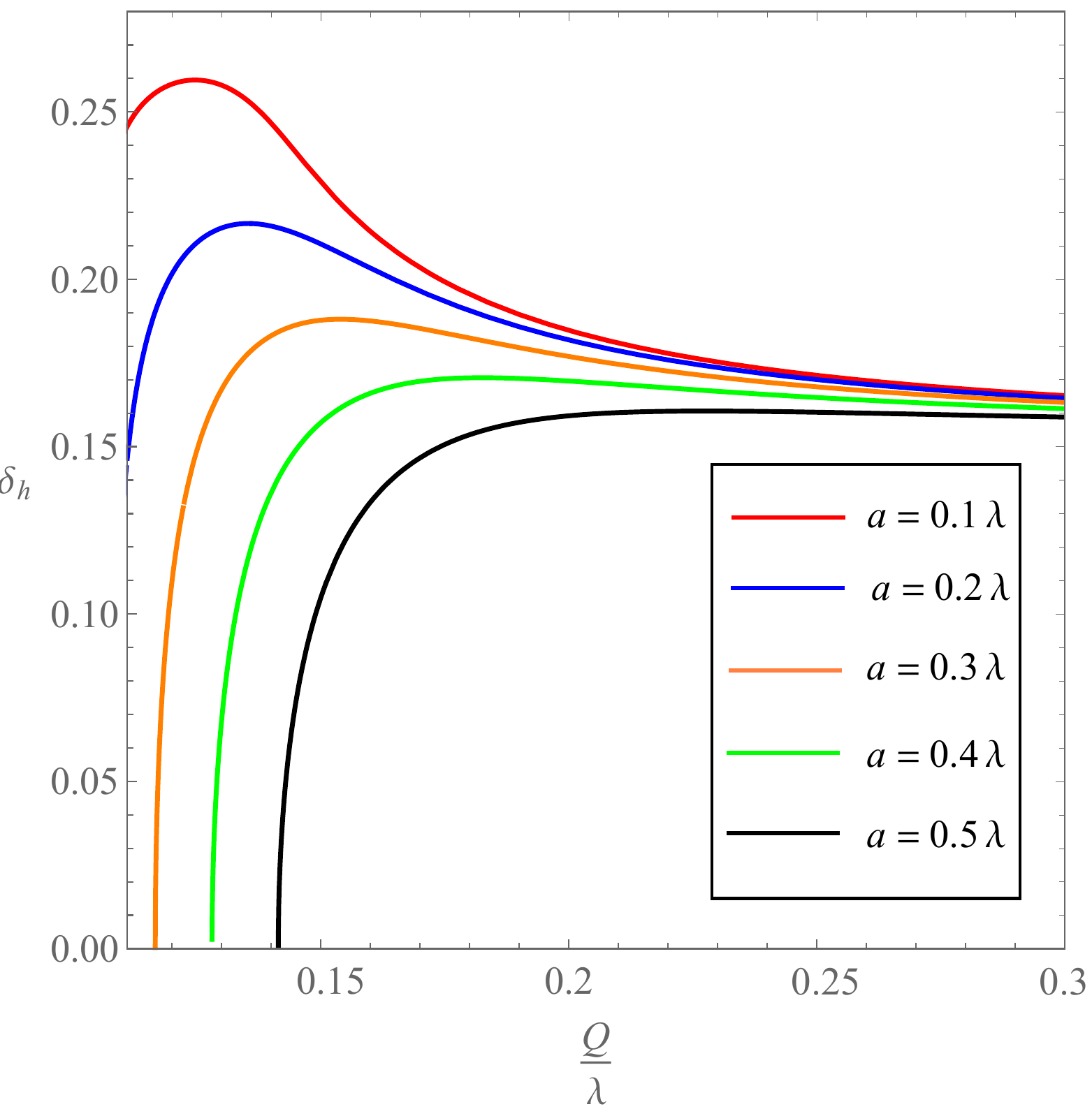}~(a)
    \includegraphics[width=8.3cm]{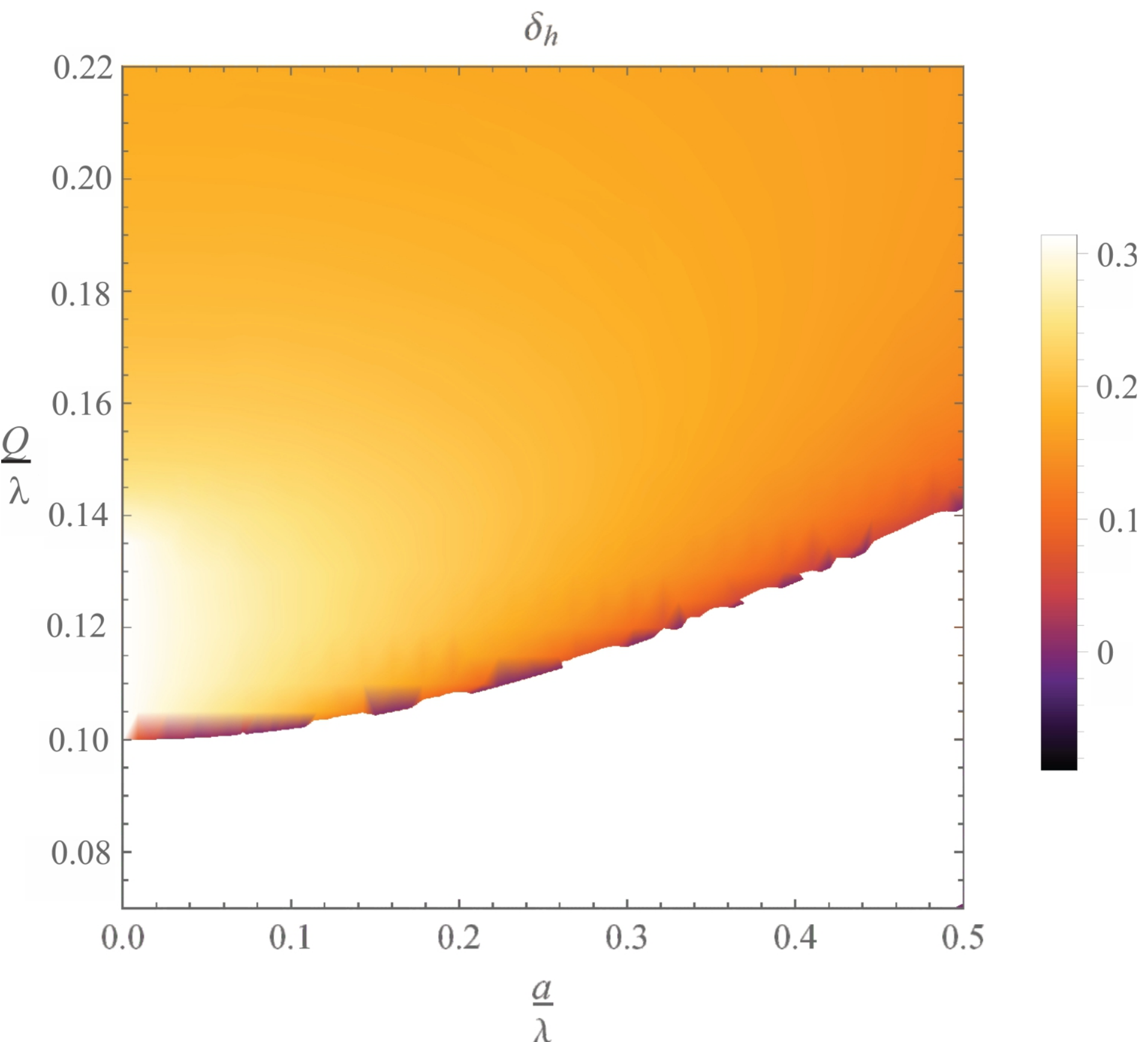}~(b)
    \includegraphics[width=7cm]{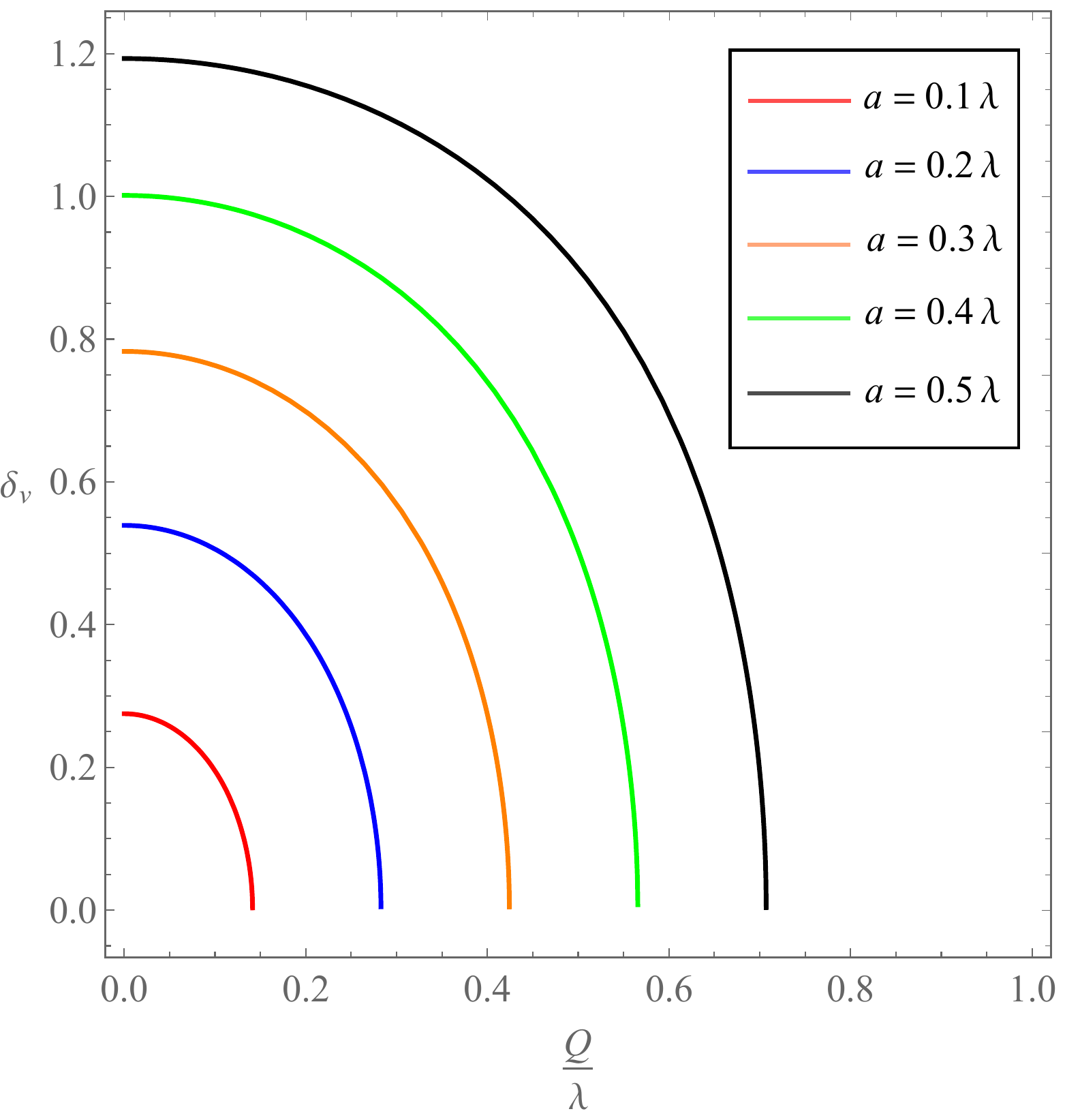}~(c)
    \includegraphics[width=8.4cm]{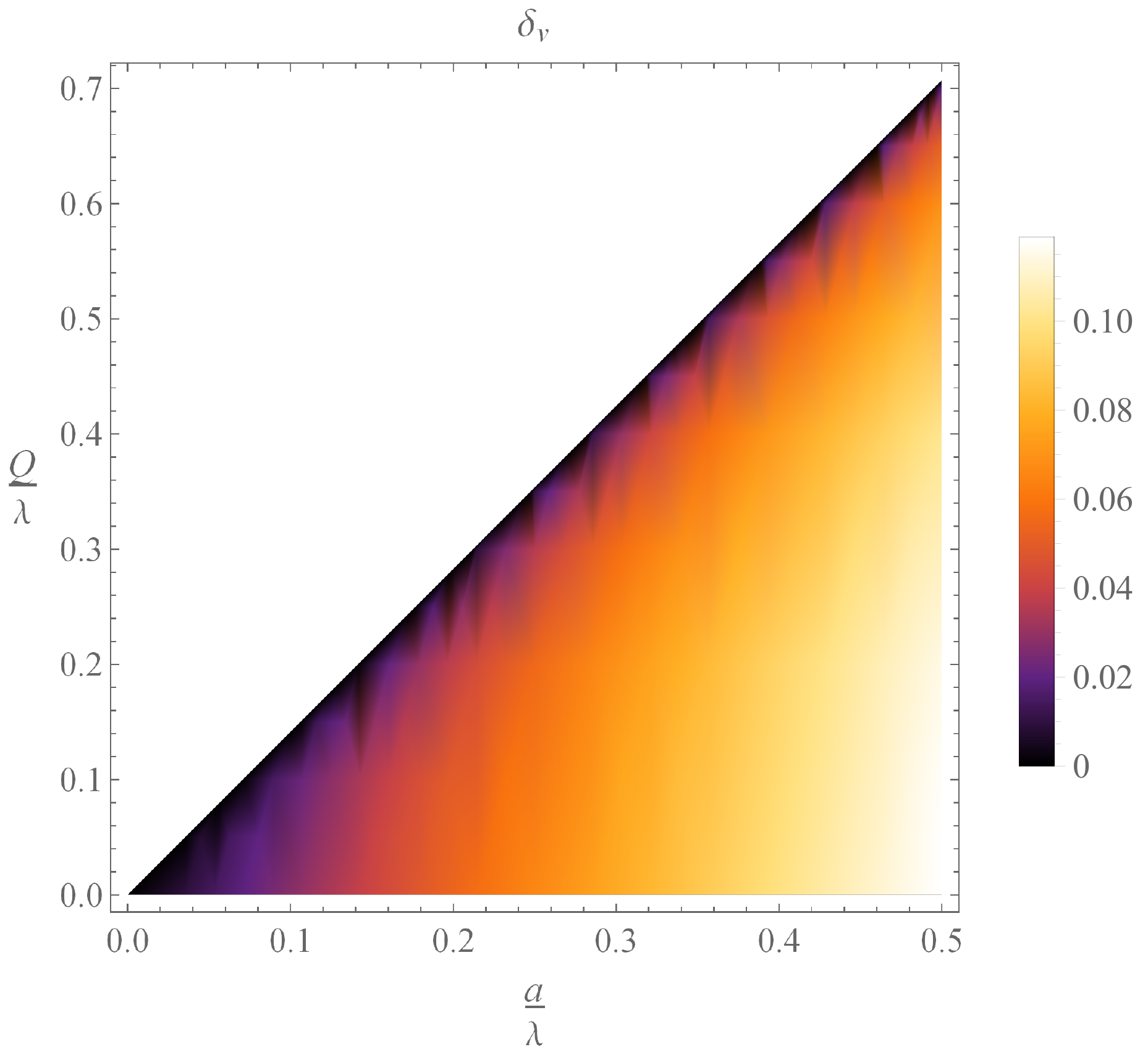}~(d)
    \includegraphics[width=7.4cm]{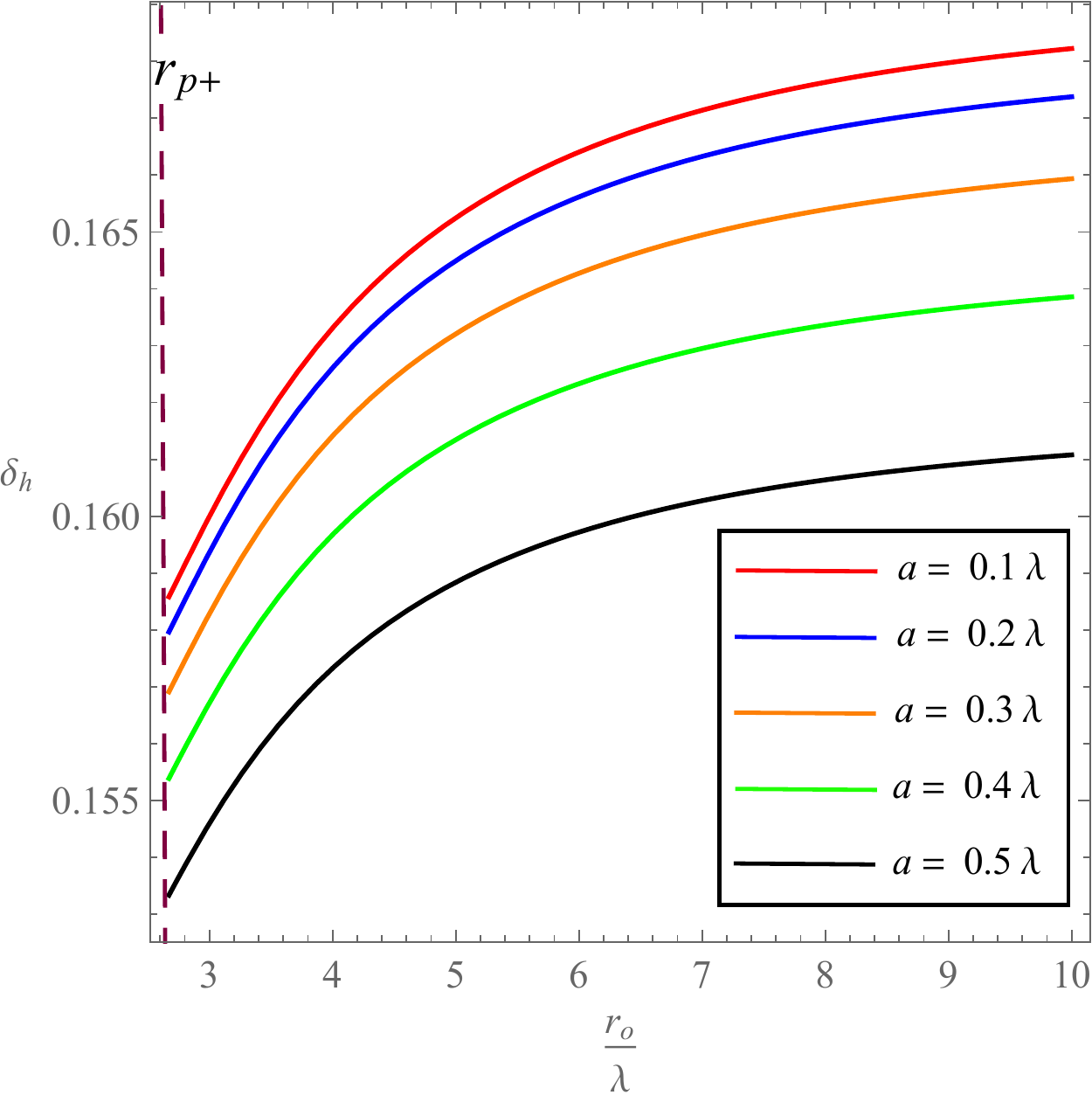}~(e)
    \includegraphics[width=7.4cm]{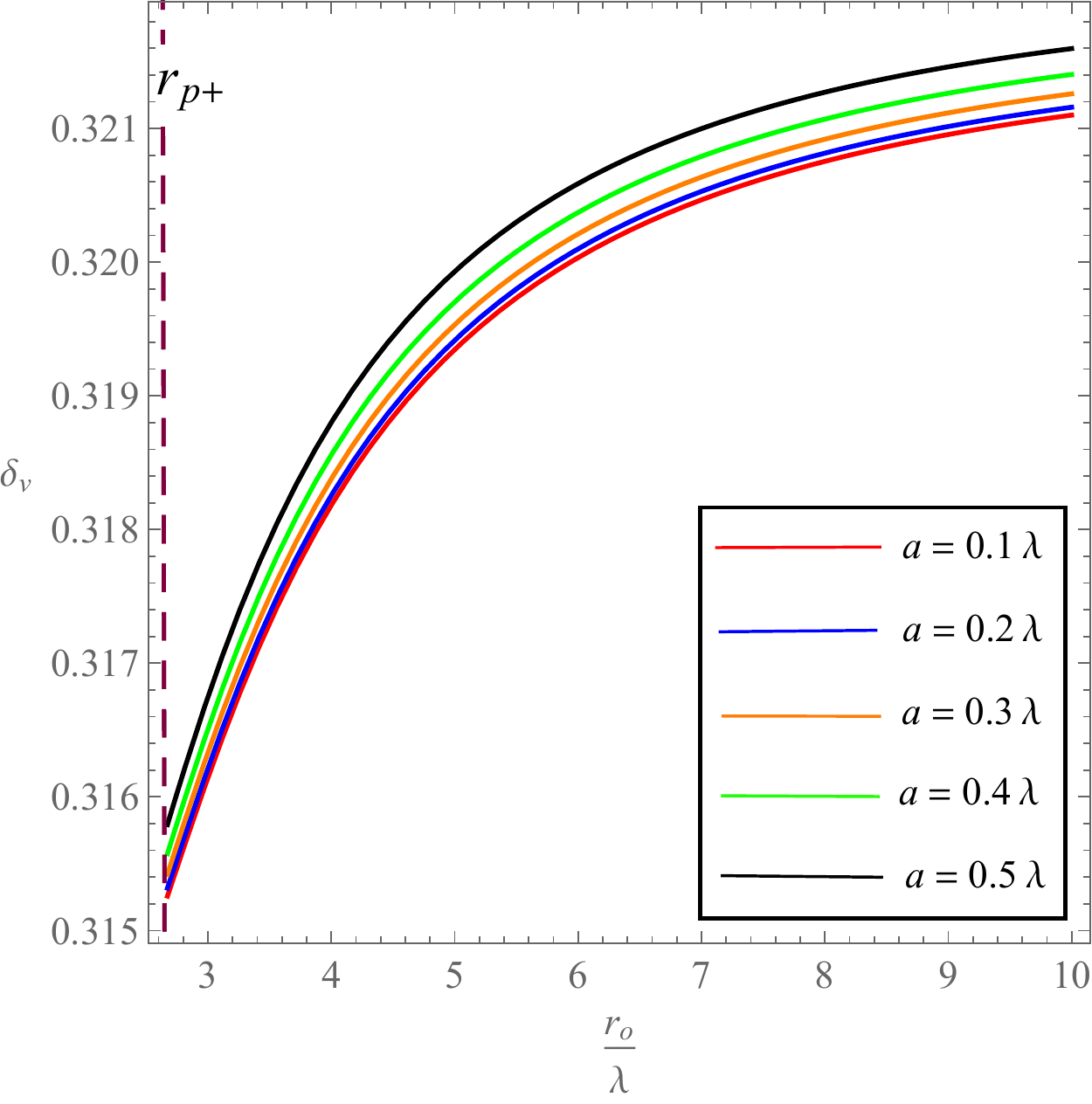}~(f)
    \caption{{The variations of the horizontal (diagrams (a), (b)) and the vertical (diagrams (c), (d)) angular diameters of the RCWBH as functions of $Q$ and $a$, plotted for $\lambda=10$ and $r_o=5\lambda$. In the left panels, the changes of angular diameters in terms of $Q$, have been plotted distinctly for five values of $a$. In the right panels, $Q$ and $a$ have been let to change freely. The bottom panel diagrams ((e), (f)), correspond respectively to the dependence of the angular diameters on the variations of the observer's location $r_o$, which have been plotted for $Q = 0.3 \lambda$. }}
    \label{fig:deltah,deltav}
\end{figure}
As observed from the diagrams, each of the horizontal diameter curves has a maximum, which is more significant for smaller $a$. Furthermore, the horizontal size is larger for smaller $a$, and all the curves tend to a same value as $Q$ increases. 
On the other hand, the curves corresponding to the vertical
diameter, expose a smooth decrease in value with respect to increase in $Q$, for each of the cases. Also, increase in the spin parameter only increases the vertical size of the shadow, and in contrast with the previous case, the
vertical size is smaller for smaller $a$. These can be inferred, as well, from the density plots in the right panel diagrams.
{Moreover, as seen in Eqs. \eqref{eq:sinh} and \eqref{eq:sinv}, the other effective factor in the angular sizes of the shadow, is the observer's distance $r_o$. As shown in the bottom panels of Fig. \ref{fig:deltah,deltav}, by increase in $r_o$ outside $r_{p+}$, the shadow increases in size until its angular diameters reach a constant maximum value, within a finite distance inside the causal region. In fact, for the vertical diameter, the radial distance $r_o$ could encounter a minimum and a maximum, that can be obtained from Eq. \eqref{eq:C(rv)} as
\begin{equation}\label{eq:minimumrv}
r^2_{o\substack{\max\\\min}}  = 
\frac{1}{2}\left(\lambda^2\pm\sqrt{\lambda^2\left(4a^2-Q^2+\lambda^2\right)}\right),
\end{equation}
where, $r_{o\min}$ exists only for $Q>2a$, for which, $r_{p-}$ also exists. It is important to note that, the increase in size of the shadow by distance, is a consequence of the peculiar contribution of the cosmological term $\lambda$ for the RCWBH. This term becomes more dominant as $r_o$ increases. 
}

{It is of worth comparing the above vertical diameter, with the angular diameter assigned to the shadow of M87*. This way, we can obtain an estimation for the charge component of the RCWBH, if it has the same angular diameter. For M87*, located at the distance $r_{o*} \approx 5.18\times10^{23}  \mathrm{m}$ from earth, the values associated with the lapse function components in Eq.~\eqref{par1} are $\tilde{m}_* \approx 6.4\times10^9M_\odot$ and $\tilde{r}_* \approx 1.82\times10^{13} \mathrm{m}$ \cite{Akiyama:2015}. The angular diameter of the shadow has been observed as $\tilde{d}_* = (42\pm3)~\mathrm{\mu a s}$}\footnote{$1~\mathrm{\mu a s} \approx 4.85\times10^{-12}$ rad.} \cite{Akiyama:2019}. {To do the comparison, we fix the cosmological component of the spacetime to the current value of the cosmological constant, by letting $\tilde{\varepsilon}_* = \Lambda_0\approx 1.11\times10^{-52}~ \mathrm{m}^{-2}$ \cite{Planck:2016}. Furthermore, the recent evaluation of the spin parameter of M87* is $a_* = 0.9\pm0.05$ \cite{Tamburini:2019}. Assuming these values together with the equivalence  $\tilde{d}_* = \delta_v$, and by taking into account the expression in Eq.~\eqref{eq:C(rv)}, we get $Q_* \approx 1.8\times 10^{34} \mathrm{m}$. This value is approximately equivalent to $2.1\times 10^{17} \mathrm{C}$, which is nearly the charge of $10^{36}$ protons}\footnote{The change from SI to geometric units for the electric charge is $Q_{(\mathrm{Coulomb})} = \sqrt{\frac{4\pi\epsilon_0 c^4}{G}}~Q_{(\mathrm{meters})} =  1.15964\times 10^{17} Q_{(\mathrm{meters})}$, and the proton's electric charge is $q_p = 1.602\times 10^{-19} \mathrm{C}$. For the mass, $\tilde{m}_{(\mathrm{meters})} = \tilde{m}_{(\mathrm{kg})}\times\frac{G}{c^2}$, and we have considered $M_\odot = 1.989\times 10^{30}\mathrm{kg}$ \cite{Phillips:1992}.}. {This value corresponds to a charge density of $\tilde{\rho}_{Q*}\approx 8.32\times 10^{-24}~\frac{\mathrm{C}}{\mathrm{m}^3}$ for the black hole.} \\

In this section, we simulated the shadow of the RCWBH and calculated its angular size, by applying the geometric method given in Ref.~\cite{Grenzebach:2015}. {Furthermore, we used the shadow's vertical angular diameter to estimate the electric charge of the RCWBH, if it is supposed to have the same characteristics of M87*. We leave the discussion at this point and summarize the results in the next section.}

\section{Summary and conclusions}\label{sec:conlcusion}

The application of conformal invariance in alternative theories of gravitation has had a long history among relativists and has even been considered as a tool to solve the singularity issue in general relativity \cite{Mannheim:2012}. In this regard, joining conformal symmetry with general relativistic black hole spacetimes, it has been shown that it is possible to obtain singularity-free black holes in conformal gravity, that can be confined successfully within the mass and the radiation spectrum of the Schwarzschild and Kerr black holes \cite{Bambi:2017a,Bambi:2017,Zhou:2019}. Static and rotating black hole spacetimes inferred from the Weyl conformal theory of gravity, on the other hand, are mostly given in terms of the Mannheim-Kazanas axially symmetric solutions to the Bach equations \cite{Mannheim:1989,Mannheim1991} and are intended to replace the astrophysical explanations which are based on dark energy and dark matter scenarios. In this paper, however, we considered a particular solution to the Bach equations, which has been specified to a spherically symmetric massive source. The rotating counterpart of this solution was obtained from the Azreg-A\"{i}nou method. We then assessed the resultant stationary axially symmetric spacetime by calculating the variations of its horizons in terms of changes in the black hole parameters. In fact, similar to the Pleba\'{n}ski–Demia\'{n}ski class of solutions to the Einstein-Maxwell equations that are associated with a cosmological constant, the spacetime corresponding to the RCWBH contains two null hypersurfaces as the event and the cosmological horizons. This latter is the surface of infinite blue-shift and the former, coincides with the surface of infinite red-shift for the static case of a CWBH. For rotating black holes, however, the event horizon and the surface of infinite red-shift are separated. Regarding this, we calculated the position of this surface for the RCWBH and figured out the static limit of the black hole's exterior geometry, on which, the stationary and the static observers change their nature and commence corotation with the black hole. Accordingly, we demonstrated the variations of black hole's ergosphere with respect to changes in the electric charge and the spin parameter. Turning to the optical properties of the black hole, we used the first order light-like geodesic equations to calculate the light rays' impact parameters, which enabled us to determine the unstable photon orbits and the related photon regions around the black hole. The photon region corresponds to what the distant observers would observer from the black hole. This region is therefore, the ultimate limit of black hole's optical visibility and in fact, confines the black hole's shadow. To demonstrate the shadows of the RCWBH, we calculated the Cartesian celestial coordinates based on the method provided in Ref.~\cite{Grenzebach:2014}. The boundary of the shadow was shown to be both oblate and sharp towards the origin of the celestial coordinates, and for a given charge parameter, increase in the spin parameter shrinks the shadow, horizontally and vertically. This procedure is in fact, the reason of the shadow's deformation. We also demonstrated this, by calculating the angular sizes of an oblate shadow, applying three angular radii as proposed in Ref.~\cite{Grenzebach:2015}. {For the particular case of oblate shadows, we also compared the RCWBH and the KNdSBH, which indicated the larger size of the shadow of the RCWBH, respecting the black hole masses.}
{The evolution of the photon region and the shadow of the RCWBH depends rigorously on the presence of an electric charge. Based on this fact, we applied the calculations corresponding to the vertical angular diameter of the shadow, to obtain an estimation for the electric charge of M87*, if it was supposed to be a RCWBH. The results indicated that this black hole will have about $10^{36}$ free protons to fulfill its electric charge deposit, which is about $10^{-21}$ times the total protons of the Sun \cite{Phillips:1992}. This, in the realm of charged black holes, makes sense, because a black hole is not supposed to have active nuclear fusion. Therefore, the investigation presented here may help understating the optical evolution of charged black hole spacetimes that are inferred from alternative theories of gravity, as well as providing information about the impacts of electromagnetic constituents of such spacetimes.}

For future investigations, we can consider the motion of mass-less and massive test particles around the RCWBH, as well as investigating the relevant gravito-electromagnetic effects. These tasks are left for future studies.

\begin{acknowledgements}
The authors would like to thank the referees for their comments which significantly helped us improving the presentation of the paper. M. Fathi has been supported by the Agencia Nacional de Investigaci\'{o}n y Desarrollo (ANID) through DOCTORADO Grants No. 2019-21190382, and No. 2021-242210002. {J.R. Villanueva was partially supported by Centro de Astrof\'isica de Valpara\'iso (CAV).}
\end{acknowledgements}


\appendix

\section{The method of solving biquadratic equations}\label{app:Ap}

To solve the biquadratic equations of the form
\begin{equation}\label{a1}
x^4-a_1\,x^2+a_2=0,
\end{equation}
with $(a_1,a_2) > 0$ and $2\sqrt{a_2}\leq a_1$, we apply the change of variable $x = Z \sin \varphi$, and multiply both sides of the equation by a scalar $\alpha$. This provides
\begin{equation}\label{a2}
\alpha\, Z^4 \sin^4\varphi - \alpha \,a_1\, Z^2 \sin^2 \varphi + \alpha \,a_2 = 0.
\end{equation}
Comparison with the trigonometric identity
\begin{equation}\label{a3}
4 \sin^4\varphi -4 \sin^2 \varphi +  \sin^2 (2 \varphi) = 0,
\end{equation}
yields
\begin{equation}\label{a4}
\alpha\, Z^4 =4, \qquad \alpha \,a_1\, Z^2=4, \qquad  \alpha \,a_2=\sin^2 (2 \varphi).
\end{equation}
We therefore obtain $Z$ and $\varphi$ as
\begin{equation}
Z=\sqrt{a_1}, \quad \text{and} \quad \varphi_n={1\over 2} \arcsin\left( {2\sqrt{a_2}\over a}\right) +{n\pi\over 2} ,
\label{a4}
\end{equation}
where the period of the trigonometric function is $n\pi$. This way, the roots of Eq.~\eqref{a1} are obtained by letting $n = 0, 1$, which yields 
\begin{eqnarray}
 x_0&=&\sqrt{a_1}\sin\left( {1\over 2} \arcsin\left( {2\sqrt{a_2}\over a_1}\right)  \right),  \\
 x_1&=&\sqrt{a_1}\sin\left( {1\over 2} \arcsin\left( {2\sqrt{a_2}\over a_1}\right) +{\pi\over 2} \right) \nonumber\\
 &=& \sqrt{a_1}\cos\left( {1\over 2} \arcsin\left( {2\sqrt{a_2}\over a_1}\right)\right),    \\
  x_2&=&- x_0,\\
  x_3&=&-x_1.
\end{eqnarray}

\bibliography{Biblio_v1.bib}

\end{document}